\documentclass{emulateapj}

\usepackage{graphicx,longtable,hyperref}
\usepackage{natbib}

\newcommand{\arcs}{\mbox{\ensuremath{^{\prime\prime }}}}

\newcommand{\Rearth}{R$_\oplus$}
\newcommand{\Ms}{M$_\odot$}
\newcommand{\Rs}{R$_\odot$}
\newcommand{\RNum}[1]{\uppercase\expandafter{\romannumeral #1\relax}}

\newcommand{\gps}{\ensuremath{g_{\rm P1}}}
\newcommand{\rps}{\ensuremath{r_{\rm P1}}}
\newcommand{\ips}{\ensuremath{i_{\rm P1}}}

\newcommand{\PS}{\protect \hbox {Pan-STARRS1}}

\shorttitle{Search for WD Eclipses}
\shortauthors{Fulton}

\begin{document}

\title{A search for planetary eclipses of white dwarfs in the \PS\ medium-deep fields}

\author{
B. J.~Fulton\altaffilmark{1},
J. L.~Tonry\altaffilmark{1},
H. Flewelling\altaffilmark{1},
W. S. Burgett\altaffilmark{1},
K. C. Chambers\altaffilmark{1},
K. W. Hodapp\altaffilmark{1},
M. E. Huber\altaffilmark{1},
N. Kaiser\altaffilmark{1},
R. J. Wainscoat\altaffilmark{1},
C. Waters\altaffilmark{1}
}

\altaffiltext{1}{Institute for Astronomy, University of Hawaii at Manoa, Honolulu, HI 96822, USA}

\begin{abstract}
We present a search for eclipses of $\sim$1700 white dwarfs in the \PS\ medium-deep fields. Candidate eclipse events are selected
by identifying low outliers in over 4.3 million light curve measurements. We find no short-duration eclipses consistent with being caused by
a planetary size companion. This large dataset enables us to place strong constraints on the close-in planet occurrence rates around white dwarfs for planets as
small as 2 \Rearth. Our results indicate that gas giant planets orbiting just outside the Roche limit are rare, occurring around less than 0.5\% of white dwarfs.
Habitable-zone super-Earths and hot super-Earths are less abundant than
similar classes of planets around main-sequence stars. These constraints give important insight into the
ultimate fate of the large population of exoplanets orbiting main sequence stars.

\end{abstract}

\section{Introduction}
Searches for planets outside our solar system have focused primarily on hydrogen-burning main-sequence stars similar to our Sun
\cite[e.g.][]{Bakos04, Howard10b, Borucki10}. As we discovered that planets are nearly ubiquitous in our Solar neighborhood \citep{Howard10a}
and in the \emph{Kepler} field \citep{Petigura13}
searches around M-dwarfs gained popularity \citep[e.g.][]{Nutzman08}. Studies of M-dwarfs enjoy a boost in sensitivity to small planets
because transits block a larger fraction of the stellar disk and induce a larger amplitude reflex motion of the star around the barycenter due to their low mass.
Some studies have also searched for and explored the planet occurrence rates as a function of stellar mass
from M-dwarfs to intermediate-mass subgiants \citep{Johnson07}. Microlensing campaigns survey stars of many types
and are sensitive to planets around all massive hosts
regardless of their stage in stellar evolution \citep{Gaudi12} but followup characterization of these planets is impossible.
However, there have been few dedicated searches for planets around white dwarfs (WDs).

Many studies including \citet{Mullally07}, \citet{Farihi08}, and \citet{Kilic09} searched for infrared-excess indicative of planetary companions to WDs. They detected several brown dwarf companions \citep{Zuckerman92, Farihi05, Steele09} but no planetary-mass objects. \citet{Mullally07} also searched for companions using pulsations of WDs to
look for periodic deviations in the pulse arrival times caused by an orbiting companion. They find evidence of a 2.4 M$_{J}$ companion in 
a 4.6 year orbit. \citet{Hogan09}, and \citet{Debes05a} conducted high contrast imaging surveys of nearby WDs to search for low-mass
companions at large separations. \citet{Burleigh06} found a brown dwarf in the near-IR spectrum of WD 0137-349 with an orbital period of only 2 hours. This object may have survived the common-envelope phase or migrated from larger orbital distances after the formation of the WD. \citet{Faedi11} conduct a transit search for a sample of 174 WDs using SuperWASP data \citep{Pollacco06} and find no eclipsing companions but can put only weak constraints on the planet occurrence rates due to their small sample size ($<$10\% for Jupiter-size planets).
\citet{Drake10} search for eclipses of $\sim$12,000 color-selected WDs using Catalina Sky Survey photometry and Sloan Digital Sky Survey spectroscopy.
They find 20 eclipsing systems and three of them have radii consistent with substellar objects and no detectable flux in the spectra.

WDs have radii only $\sim$1\% of the Sun, or about the same size as the Earth. This implies that an Earth-sized object
transiting the WD with an impact parameter of 1.0 would cause a complete occultation. Although these occultations are short-duration, 
they can be easily detected from small ground-based telescopes with short exposure times and relatively low photometric
precision \citep{Drake10}. In addition, the most common WDs are old and cool with surface temperatures of $\sim$5000 K. Their small radii
and low surface temperatures imply that their luminosity is low, with typical values of $\sim$10$^{-4}$ L$_{\odot}$, and the habitable zone
is close-in \citep[a$\sim$0.01 AU,][]{Agol11} giving rise to significant transit probabilities. This makes Earth-size planets orbiting in the habitable
zones of old, cool WDs relatively easy to detect via the transit method.

Most main-sequence stars, including our Sun, will eventually
end their lives slowly cooling as WDs. Since approximately 50\% of main sequence stars host at least one planet \citep{Mayor11}
it is interesting to consider their fate as the star evolves into a WD.
It is unlikely that any planets inside $\sim$1 AU would survive engulfment by their host stars as they
expand onto the red giant branch but it is unclear what becomes of the planetary debris. Since WDs quietly cool for the age of the universe, it is conceivable that new planets
could form out of the debris of a previous generation of planets. Migration of planets from outside of 1 AU is also plausible, but little theoretical work has been
done on the formation or migration of planets hosted by WDs. Several studies have identified pollution by
heavy elements on the surfaces of WDs \citep{Zuckerman10} and IR excess indicative of a debris disk \citep{Debes11}. Extensive work has been done
to identify the chemical composition of this pollution. Silicates and glasses were detected in the atmosphere of six WDs by \citet{Jura09} and interpreted
as signs of accretion of asteroid-like bodies onto the WD. A detailed study by \citet{Xu14} using data from the Keck and Hubble Space Telescopes showed
strong evidence that the composition of metals in the atmospheres of WDs G29-38 and GD 133 closely mirror the composition of the bulk Earth. Furthering the idea that close-in terrestrial planets orbit and eventually accrete onto WDs.

We present a systematic search for eclipses of WDs by planetary-size objects in the \PS\ medium-deep fields \citep{Tonry12}.
We use a combination of astrometric and photometric selection techniques to identify 3179 WDs with a range of ages and temperatures.
Each WD was observed on 1000-3000 epochs during the past 5 years for a total of 4.3 million measurements. Although we do not detect any substellar companions,
this large number of observations allows us to place tight constraints on the occurrence rates of planets orbiting WDs.


\section{Methods}
\subsection{WD sample}
We analyze a total of 3179 WD candidates spread across the 10 medium-deep fields spanning 70 square degrees on the sky. Each field is observed on 1000-3000 epochs with four to eight consecutive 240 s exposures per night. Our sample of WDs is segregated into two categories.
We identify 661 targets using their proper motions as described in \citet{Tonry12} (astrometric sample hereafter).
These objects have a high probability of being bona fide WDs and a very low contamination rate.

\begin{figure}[h]
\epsscale{1.25}
\plotone{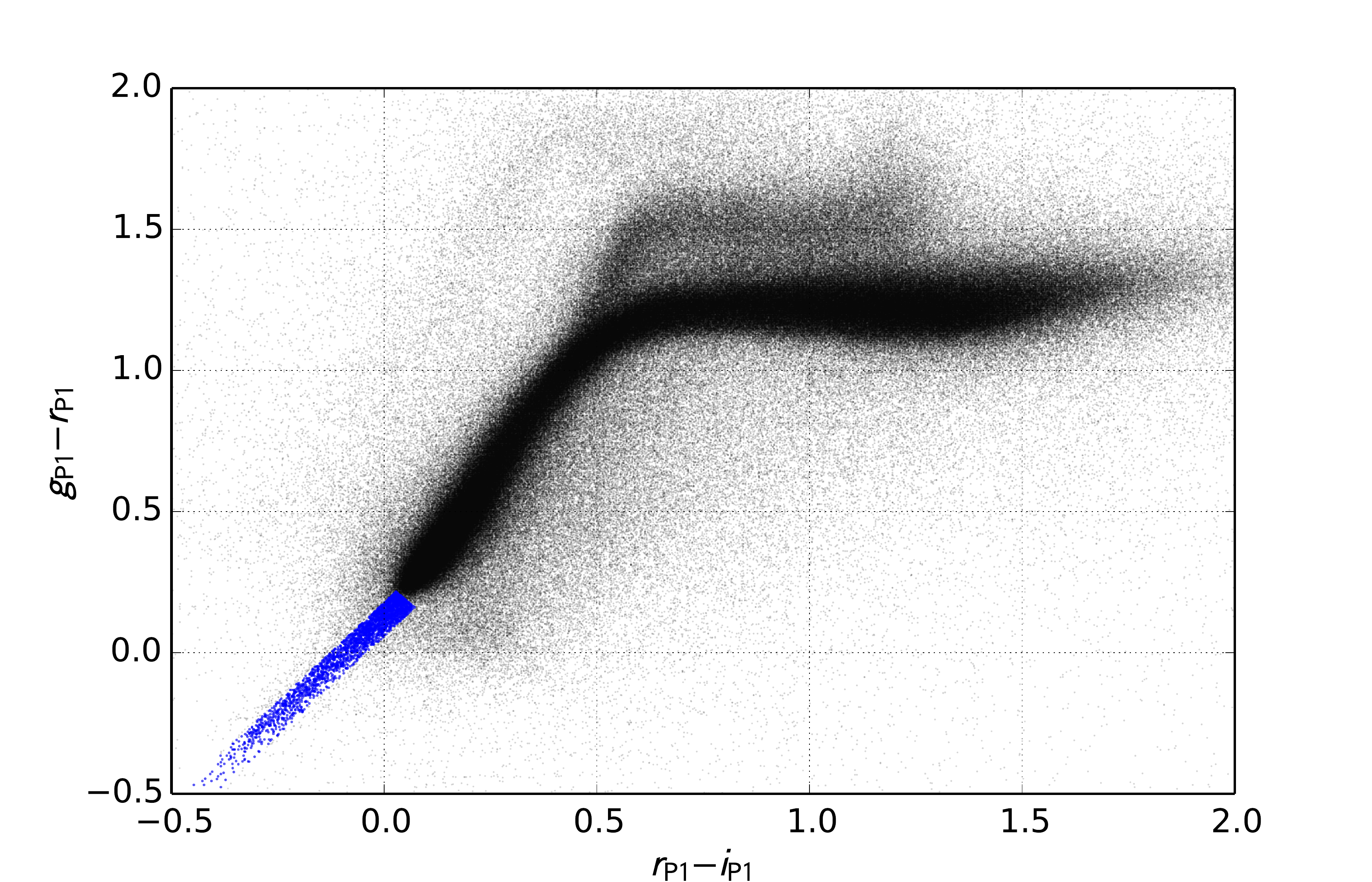}
\centering
\caption{The color-selected WDs (blue points) are identified by the narrow tail of extremely blue stars in the (\gps-\rps) vs. (\rps-\ips) color plane. The small black points are all detections from the deep stacks not selected for either the control or WD samples.}
\label{fig:color}
\end{figure}

The remaining 2518 WDs were selected based on their photometric colors (color-selected sample hereafter). We use the following criteria to select the locus of hot, blue stars from the (\gps-\rps) vs. (\rps-\ips) color plane shown in Figure \ref{fig:color}; $(\gps-\rps) < 0.18 + 1.4(\rps-\ips)$, $(\gps-\rps) > 0.06 + 1.4(\rps-\ips)$, $(\gps-\rps) < 0.25 - 1.25(\rps-\ips)$, and $\ips<22$. This sample is restricted to hot WDs
due to the requirement of blue colors and is likely contaminated by other hot stars.

To quantify the contamination rate of the color-selected sample we created a Besancon galactic simulation of the medium-deep fields \citep{Robin03}. When we make the same color-cuts we find that 42\% of the stars are bonafide WDs according to the model. The stars that are within this locus but not WDs are mostly distant A and B-type subdwarfs in the halo of the galaxy. Closer F-type subdwarfs would also fall into the locus, but are mostly far too bright to be included in our sample. We also find that the contamination rate is highly dependent on appearant magnitude with the fainter stars being much more likely to be WDs. We assume a 58\% contamination rate for our color-selected sample for all further analysis. This reduces our total number of WDs to 1718.

\subsection{Control sample}
Our control sample consists of stars with similar magnitudes and colors to the astrometrically-selected WDs but with undetectable proper motions. These should be
relatively hot stars with radii much larger than WDs around which we would not expect to see the very short-duration
eclipses indicative of a planet occulting a WD. We can compare the number of potential eclipses found in the WD sample to
the number that we find in the control sample to better understand the frequency of eclipse-like events
caused by non-astrophysical effects.

\begin{figure}[h]
\epsscale{1.25}
\plotone{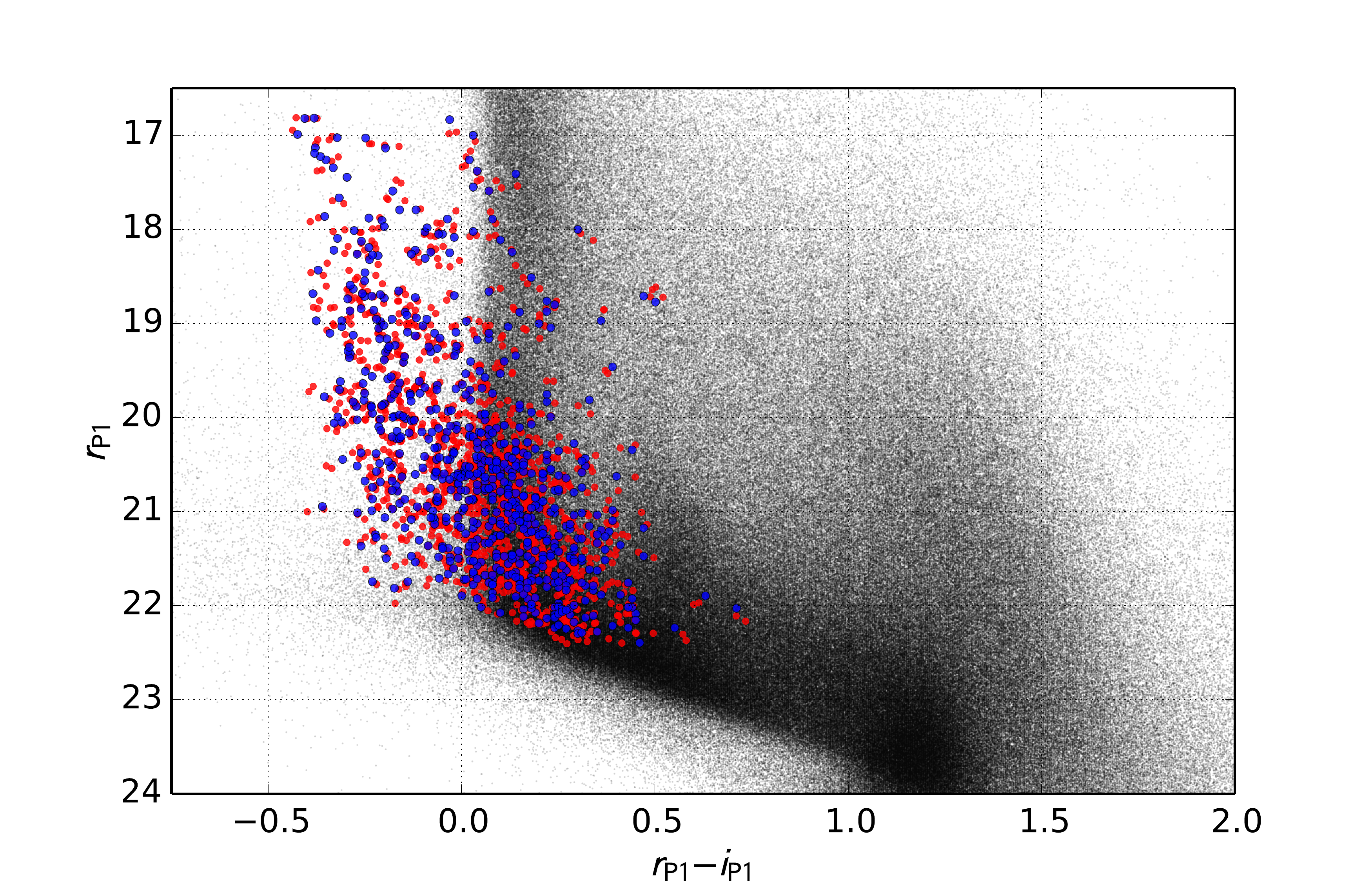}
\centering
\caption{Astrometrically-selected WDs (blue) and control sample stars (red). The small black points are all detections from the deep stacks
that were not selected for either the control or WD samples.}
\label{fig:control}
\end{figure}

We select the control sample by binning the astrometric sample of WDs in 2-dimensional color bins of $\rps$ vs. $(\rps-\ips)$. For each bin that contains
at least one WD we select two times the number of WDs in that bin from a sample of all stellar detections derived from
deep stacks of the medium-deep fields excluding stars that are already part of the WD samples. Figure \ref{fig:control} shows our
control sample and astrometric WD sample in the $\rps$ vs. $(\rps-\ips)$ color plane.
If fewer than three field stars are available in a particular bin we select all available stars. This produces a total of 1296 stars for the control
sample which is later trimmed down to 1288 by removing RR-Lyrae, Delta-Scuti and other variable stars (see \S\ \ref{sec:det}).

\subsection{Light curves}
Light curves are extracted for each WD and control sample star by directly analyzing the first-level \PS\ photometry product (SMF files).
These SMF files consist of the raw photometry extracted from the calibrated images before a zero-point or precise world coordinate system (WCS)
is established. Each camera exposure corresponds to a single SMF file. For each SMF file we first find the WCS solution
in order to associate pixel locations with sky positions. We then associate the per-image detections with detections
in deep stacks for each field and extract the PSF-fitted photometry to obtain raw instrumental magnitudes. We fit for the photometric
zero-point using the technique described in \citep{Schlafly12}. The instrumental
magnitudes for all detections within 5 arcminutes of the target are also extracted
and recorded along with the target instrumental magnitudes.
All epochs for which a target could not be matched to a detection in the SMF file are carefully recorded and the neighboring star photometry
is still extracted if available. This ensures that we are sensitive to large decreases in flux that may cause the target to fall below the detection
threshold in a particular image and in some cases we can use the photometric statistics of the neighboring stars to explain the non-detection.
We also record the pixel locations relative to the entire CCD array and particular chip for each epoch.

\subsection{Eclipse detection}
\label{sec:det}
\begin{figure}[h]
\epsscale{1.2}
\plotone{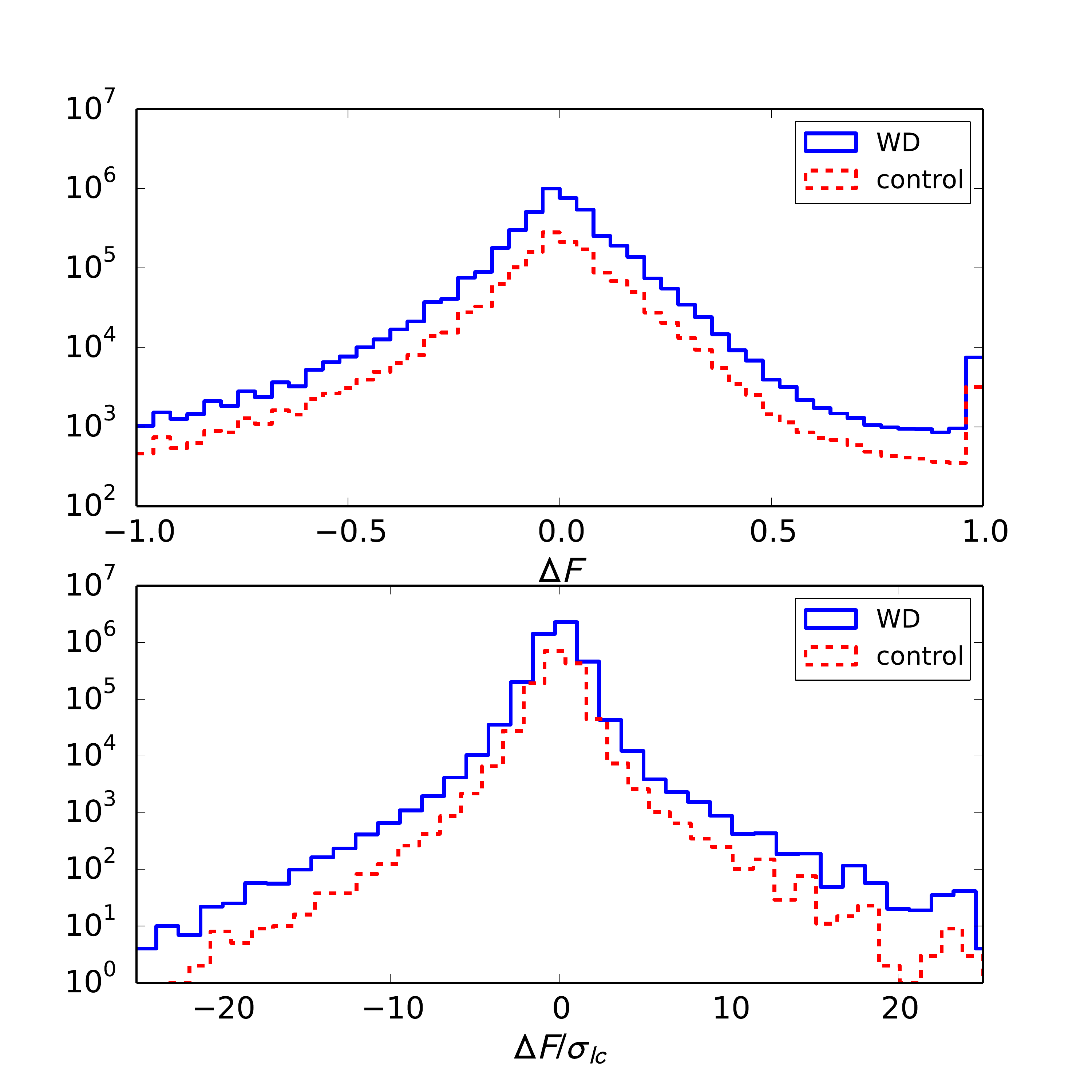}
\centering
\caption{\emph{Top:} Distribution of relative flux measurements for all WD and control sample stars. The solid blue line is the distribution for the WDs and the dashed red line is for the control sample stars.
\emph{Bottom:} Distribution of relative flux measurements divided by the measurement uncertainties corrected by adding in quadrature
the reported measurement uncertainties with the standard deviation of the light curve (by filter). Measurements with $\Delta F/\sigma_{\rm lc}\geq 5$
are considered eclipse candidates. As in the top panel, the solid blue line is the distribution for the WDs and the dashed red line comes from the control sample stars.}
\label{fig:sighist}
\end{figure}

Since eclipses are rare and extremely short duration traditional periodic search algorithms such as the box-least-squares periodogram \citep[BLS,][]{Kovacs02} fail to recover such signals. BLS excels at detecting signals in the regime of many transits with low single-event S/N but planetary eclipses of our target stars would produce very infrequent, but very deep, high S/N eclipses. Instead we employ an extremely simple eclipse detection technique. We look for low outliers
in the light curves (dropouts) that are caused either by a complete non-detection or show a deficit of flux relative to the median flux level ($\Delta F$) that is greater than five times the measurement uncertainty ($\Delta F/\sigma_{\rm lc} \geq 5$). Figure \ref{fig:sighist} shows the distribution of $\Delta F$ and $\Delta F/\sigma_{\rm lc}$ for all light curves.

\begin{figure}[ht]
\epsscale{1.15}
\plotone{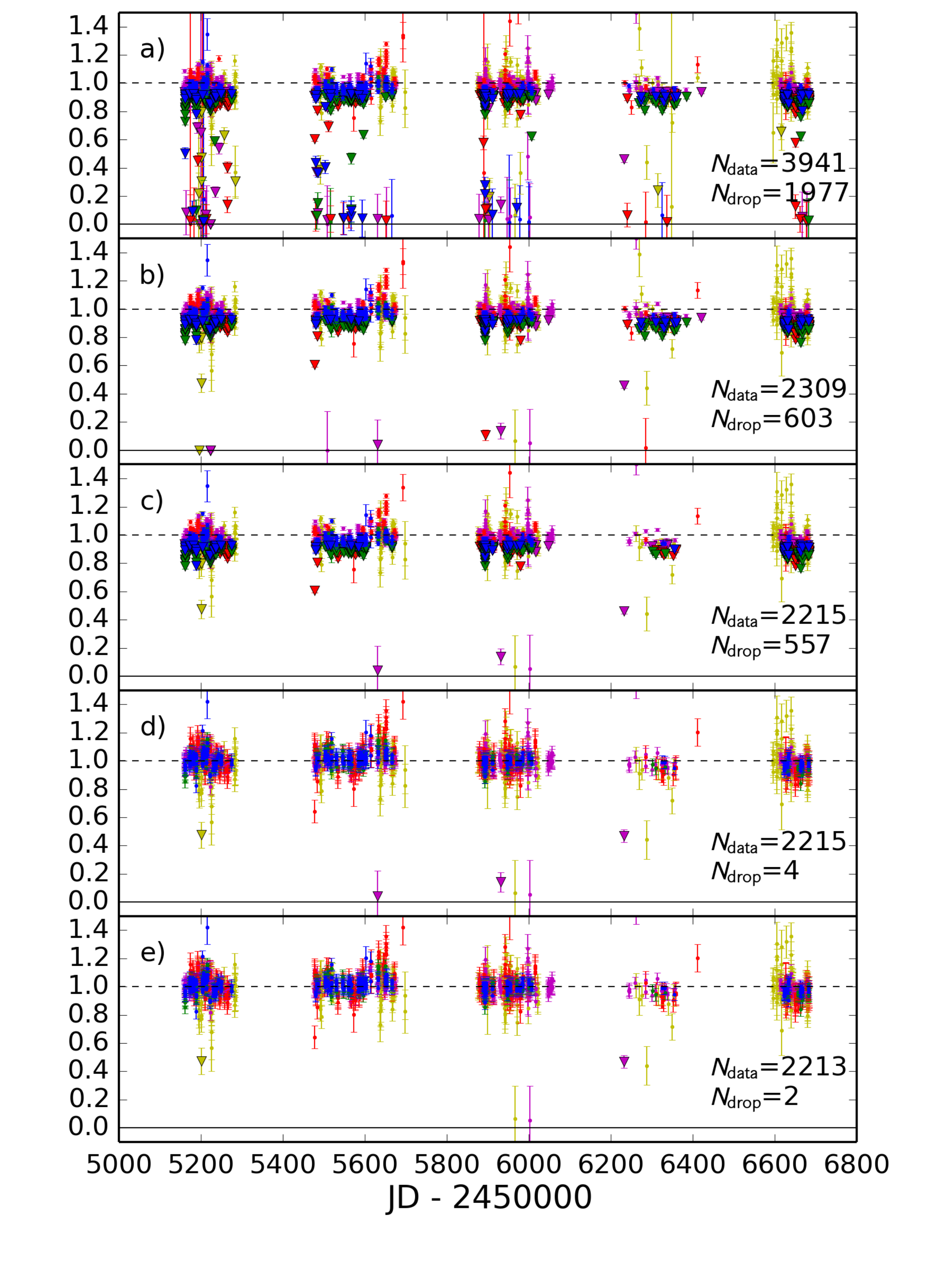}
\centering
\caption{Illustration of the filtering process for a light curve of a typical g'=18.9 WD in medium-deep field 3. The total number of measurements and the number of dropouts ($\Delta F/\sigma_{\rm lc}\geq 5$) are shown the the lower-right of each panel. Dropout candidates are plotted as triangles.
\emph{a)} Raw light curve before any filtering. Error bars are equivalent to the reported measurement uncertainties. Notice the large number (1977) of dropout candidates.
\emph{b)} Light curve after applying the chip location-based filters described in \S \ref{sec:det}.
\emph{c)} Light curve after removing measurements in which neighboring stars show large deviations from the median flux level or large scatter.
\emph{d)} Light curve after de-correlating against the neighboring star relative flux and re-scaling the measurement uncertainties by adding the reported uncertainties in quadrature with the standard deviation of the light curve in each filter. This tends to inflate the error bars and pushes the vast majority of dropout
events below the 5-sigma cutoff.
\emph{e)} Light curve after the final level of photometry-based filtering. In this stage we compare the CCD pixel positions of the stars during dropout events
with known masked regions of the CCD array. Two dropout events remain after all photometry-based filters. Postage stamp images are downloaded and visually inspected for the remaining dropout events.
}
\label{fig:lcfilters}
\end{figure}

\begin{figure*}[ht]
\epsscale{1.15}
\plotone{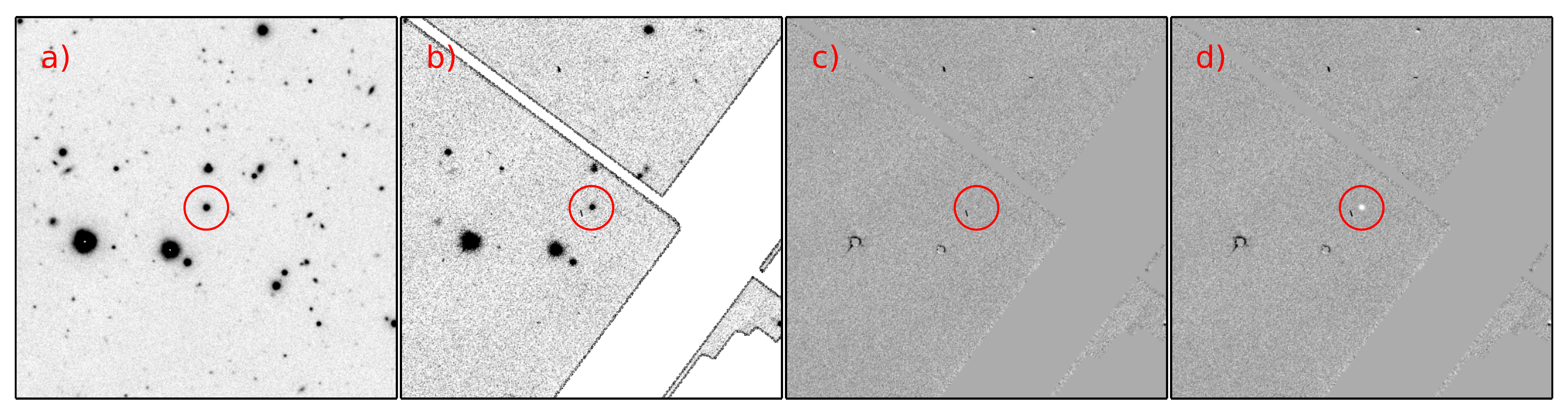}
\centering
\caption{Candidate z-band eclipse with a reported depth of 53\% that was not filtered by the automated filtering techniques described in \S \ref{sec:det}. \emph{a)} 5.8 hour stack of the $2\farcm5\times2\farcm5$ region centered on the target. The target WD is circled. \emph{b)} The same field of view as panel \emph{a} from the single exposure corresponding to the reported 53\% deep eclipse. \emph{c)} A difference image of the stack in panel \emph{a} convolved and scaled to match the PSF and subtracted from the dropout image in panel \emph{b}. Notice that all stars -- including the target -- show no detectable residual flux. \emph{d)} Same as panel \emph{c} with a synthetic 53\% eclipse injected onto the target before the image subtraction. The negative residuals on the target are clearly evident.
}
\label{fig:candidate}
\end{figure*}

\begin{figure*}[ht]
\begin{equation}
\label{uniformoccult}
1-F(p,b) = 
\left\{\begin{array}{ll}
0  &  1+p < b \\
\frac{1}{\pi} \left[p^2 \kappa_0+\kappa_1-\sqrt{{4b^2-(1+b^2-p^2)^2\over 4}}\right] &  |1-p| < b \le 1+p \\
p^2 & b \le 1-p\\
1  &  b \le p-1,\\
\end{array}\right.
\end{equation}

\begin{equation}
b(t) \approx a\sqrt{\sin^2\left(\Omega + \omega t + \alpha_0\right) + \sin^2{\theta}\cos^2\left(\Omega + \omega t + \alpha_0\right)}
\end{equation}

\end{figure*}

\begin{deluxetable*}{l r r r r}
\centering
\tablenum{1}
\tabletypesize{\tiny}
\tablehead{
\colhead{} & \multicolumn{2}{c}{WD} & \multicolumn{2}{c}{Control} \\
\colhead{Filter} & \colhead{N$_{\rm detections}$} & \colhead{N$_{\rm dropouts}$} & \colhead{N$_{\rm detections}$} & \colhead{N$_{\rm dropouts}$}
}
\startdata
No filters & 5,650,109 & 6,963,603 & 1,814,296 & 3,106,873 \\ 
CCD location-based filters & 4,757,706 & 1,771,860 & 1,523,212 & 622,154 \\ 
Neighboring star filter & 4,509,855 & 1,651,266 & 1,439,106 & 577,904 \\
Re-calculate measurement errors & 4,349,232 & 15,120 & 1,363,979 & 3,983 \\
Remove masked CCD regions & 4,343,011 & 9,000 & 1,362,535 & 2,570 \\
\enddata
\label{tab:detections}
\end{deluxetable*}

The raw light curves are heavily contaminated with non-detections and large flux drops that could be indicative of an eclipse event
or a variety of non-astrophysical scenarios.
For every dropout we first check that the star did not fall off of, or too near the edge
of a chip. We initially noticed that the dropout events were concentrated around the edges of the chips. This is likely caused by
the PSF fit failing due to a strong gradient in the background region near the edges of the chips. This effect is worse at the corners of the chips
near the readout electronics. For these reasons we remove all light curve measurements that fall within 10 pixels (2$\farcs$5) of an
edge or within 100 pixels (25$\arcs$) of a corner. We consider this filter unbiased with respect to eclipses because there is no reason to expect that real eclipses would preferentially occur when the stars fall near the edge of a chip. Measurements with reported positions that fall between chip gaps or off the array are also excluded
at this stage. All non-detections are removed with these chip location-based filters.

If the photometry of the neighboring stars also show a large decrease in flux at the same time of the target dropout, clouds or poor seeing
is likely to blame. We exclude all measurements for which the median magnitude of the neighboring stars drops by more than 0.5 magnitudes 
or the standard deviation of the neighbor magnitudes is greater than one. We also de-correlate the target relative flux measurements
against the median $\Delta F$ of the neighboring stars to reduce the effect of spatially-dependent extinction.

Now that we have removed most of the egregious outliers from the light curve we re-define the measurement errors. We sum in quadrature the reported measurement uncertainties with the median absolute deviation (MAD) of the full light curve in each filter.
This processes always inflates the errors relative to the original measurement uncertainties and effectively removes
many remaining candidate dropouts by decreasing the value of $\Delta F/\sigma_{\rm lc}$.

At this stage we use the VARTOOLS package to create BLS and analysis-of-variance (AoV) periodograms \citep{Hartman08, Kovacs02, Schwarzenberg-Czerny89, Devor05} for all WD and control sample stars. We visually inspect these periodograms and the
light curves phase-folded to the ephemeris that corresponds to the highest peak in each periodogram. Obvious periodic variable stars are removed from further analysis. Thirty-three RR-Lyrae and Delta-Scuti stars,
one dwarf nova (IY Uma) and three variables of unknown type are identified and removed at this stage.

For the remaining dropouts we check their CCD locations against the regions of the array that are consistently masked by the \PS\ Image Processing Pipeline (IPP).
After applying all of the photometry-based tests we are left with 11570 potential dropout events and of a total of 4.3 million detections.
2570 of the dropout candidates
are from the control sample and the remaining 9000 are from the merged WD samples. This photometric filtering process for a single representative case is illustrated in Figure \ref{fig:lcfilters}, and the total number of detections and non-detections removed at each stage in the filtering process are listed in Table \ref{tab:detections}.

We download the corresponding postage stamp images for any dropouts that make it through all of these light curve-based tests for additional screening.
In addition to the postage stamp corresponding to the dropout we also download a deep stack around the target
and the image that corresponds to the light curve measurement that is closest to the median value for that filter.
We apply a few more automated filters before visually inspecting the remaining candidates.
The images are automatically inspected for masking or CCD defects around the target that
produce not-a-number (nan) values, very poor seeing, or clouds as indicated by a low zeropoint magnitude. We also perform aperture photometry on the three images
and correct to an absolute apparent magnitude using the zeropoint magnitude provided in the image headers.
Our photometry acts as check that the magnitude value reported by the IPP is in rough agreement with simple aperture photometry.

As a final step we use the
HOTPANTS implementation of the ISIS image subtraction software \citep{Alard00} to produce a difference
image using the deep stack as a template. We convolve the template to match the PSF and zero point of the dropout candidate image and subtract the convolved template from the candidate postage stamp. This difference image was used to aid the visual inspection of the 133 dropout events that could not be
explained by any of the photometry or image-based filters. Figure \ref{fig:candidate} shows an example dropout candidate image and the image-differencing processes used for visual inspection. We find no eclipse with a duration compatible with an eclipse by a substellar object in any WD or control sample light curve.


\section{Analysis}
\subsection{Theoretical eclipse probabilities}
\label{eclipseprob}

In order to asses the likelihood that an occultation would have occurred during our observing window, we calculate the probability
of eclipse as a function of eclipse depth and then apply the noise properties and eclipse detection techniques that we used in our search.
This tells us the number of eclipses we should have been able to detect as a function planet radius, orbital semi-major axis and 
the occurrence rate of planets around WDs ($\eta$).

The flux when a dark sphere eclipses a uniformly illuminated sphere is given by Equation 1 \citep{Mandel02}. Where $\kappa_1=\cos^{-1}[(1-p^2+b^2)/2b],$ $\kappa_0=\cos^{-1}[(p^2+b^2-1)/2pb]$, 
and $p \equiv R_p/R_{WD}$ is the planet to white dwarf radius ratio.

Equation 2 for $b(t)$ gives the sky-projected center to center distance between the star and planet as a function of time ($t$). $\Omega$ is the longitude of the ascending node of the planet's orbit,
$a$ is the semi-major axis of the orbit, $\omega$ is the angular frequency of the orbit, $\theta$ is the inclination of the planet's orbit,
and $\alpha_0$ is the phase of inferior conjunction. Minimizing Equation 2 leads to the smallest sky projected separation over the orbit,
$b_0 = R_{WD}\cos{\theta}$.

In order to determine the likelihood that a particular $\Delta F$ could be caused by an eclipse of the WD
we calculate the probability of eclipses as a function of eclipse depth. First, we make some assumptions for physical parameters
that are mostly constant within the parameter region of interest. We assume that M$_{WD}$ = 0.6 \Ms, R$_{WD}$ = 0.01 \Rs, all theoretical companions
are on circular orbits, no limb darkening, and 240 s as the integration time for every exposure. The probability of measuring an eclipse depth $<\Delta F(p,b)>$ at time $t$ averaged over an exposure time of $\Delta t$ is

\begin{equation}
<\Delta F(p,b)> = \frac{1}{\Delta t} \int_{t_0}^{t0 + \Delta t} F(p,b)dt
\label{eqn:deltaF}
\end{equation}

Eclipses will only occur if $|b_0| < 1 + p$, therefore the probability that a randomly-oriented, circular orbit will eclipse is

\begin{equation}
P_{\rm eclipse} = \frac{R_p + R_{WD}}{a}.
\label{eqn:eclipseprob}
\end{equation}
Although systems with $|b_0| < 1 + p$ will eclipse at some time during the orbit, the fraction of orbital phase covered during eclipse is small. The probability that any part of an
eclipse will overlap with the integration time of our survey is
\begin{equation}
P_{\rm phase} = \frac{T_{\rm dur} + E}{P},
\label{eqn:phaseprob}
\end{equation}
where $P$ is the orbital period and $T_{\rm dur}$ is the eclipse duration and $E$ is the integration time.

\begin{figure*}[ht]
\epsscale{1.15}
\plottwo{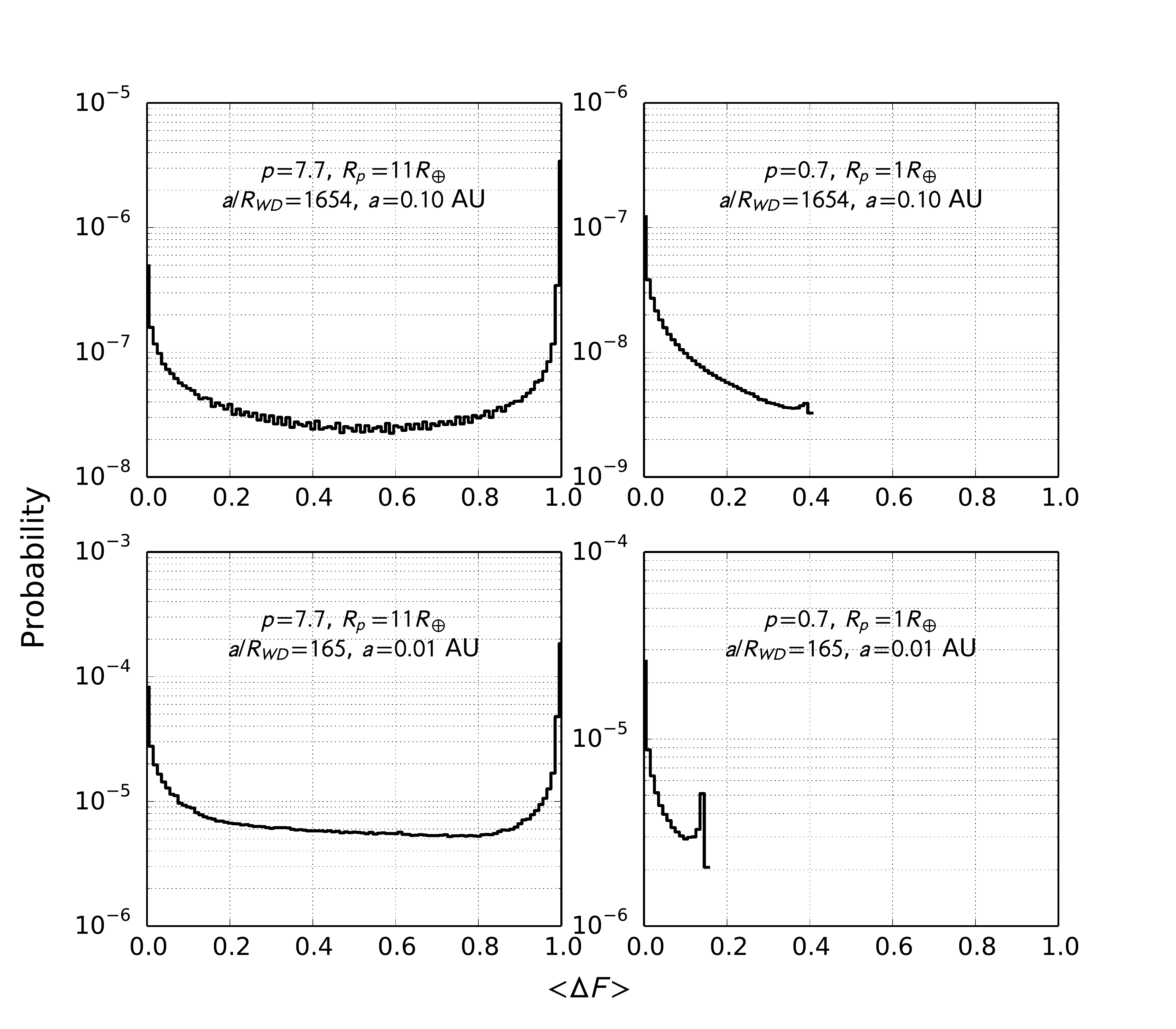}{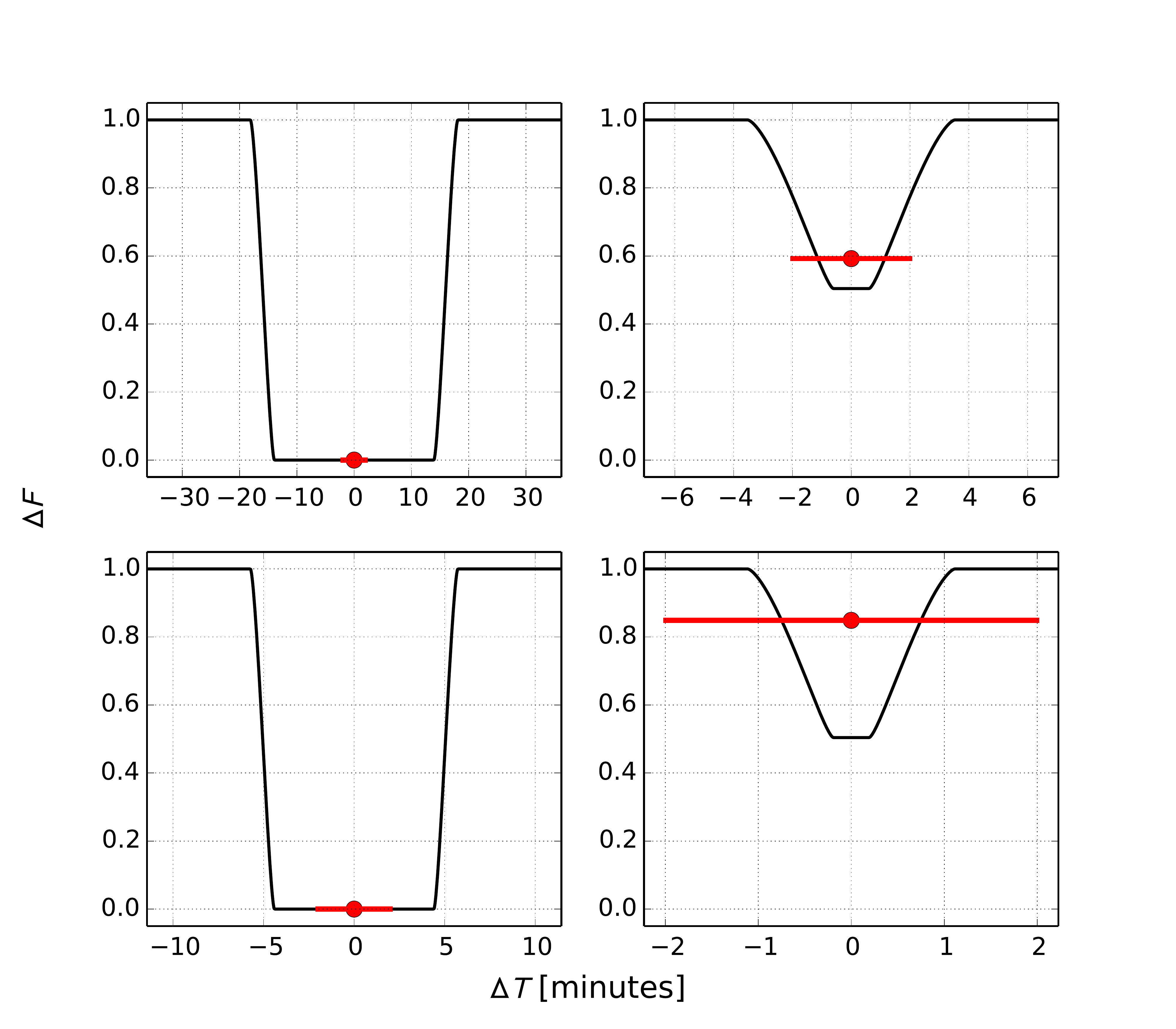}
\centering
\caption{\emph{Left:} probability of measuring an eclipse with depth $\Delta F$ during a single 240 s exposure of a random WD that hosts a
single companion with the orbital parameters shown. $p$ is the planet to star radius ratio, $R_p$ is the radius of the planet in Earth radii, $a/R_{WD}$
is the orbital semi-major axis scaled to the radius of the WD, and $a$ is the semi-major axis in AU.
\emph{Right:} Model eclipse light curves for the planet parameters shown on the left panel and an impact parameter 1.0. The red circle is the mean flux
for an exposure centered on the mid-eclipse time. The bar extending from the red circle shows the length of the exposure time. This is the largest signal that we could expect to find for planets with these parameters. This corresponds to the maximum $<\Delta F>$ bin with a probability greater than zero in the left panel.
}
\label{fig:depthprob}
\end{figure*}

\begin{figure}[h]
\epsscale{1.25}
\plotone{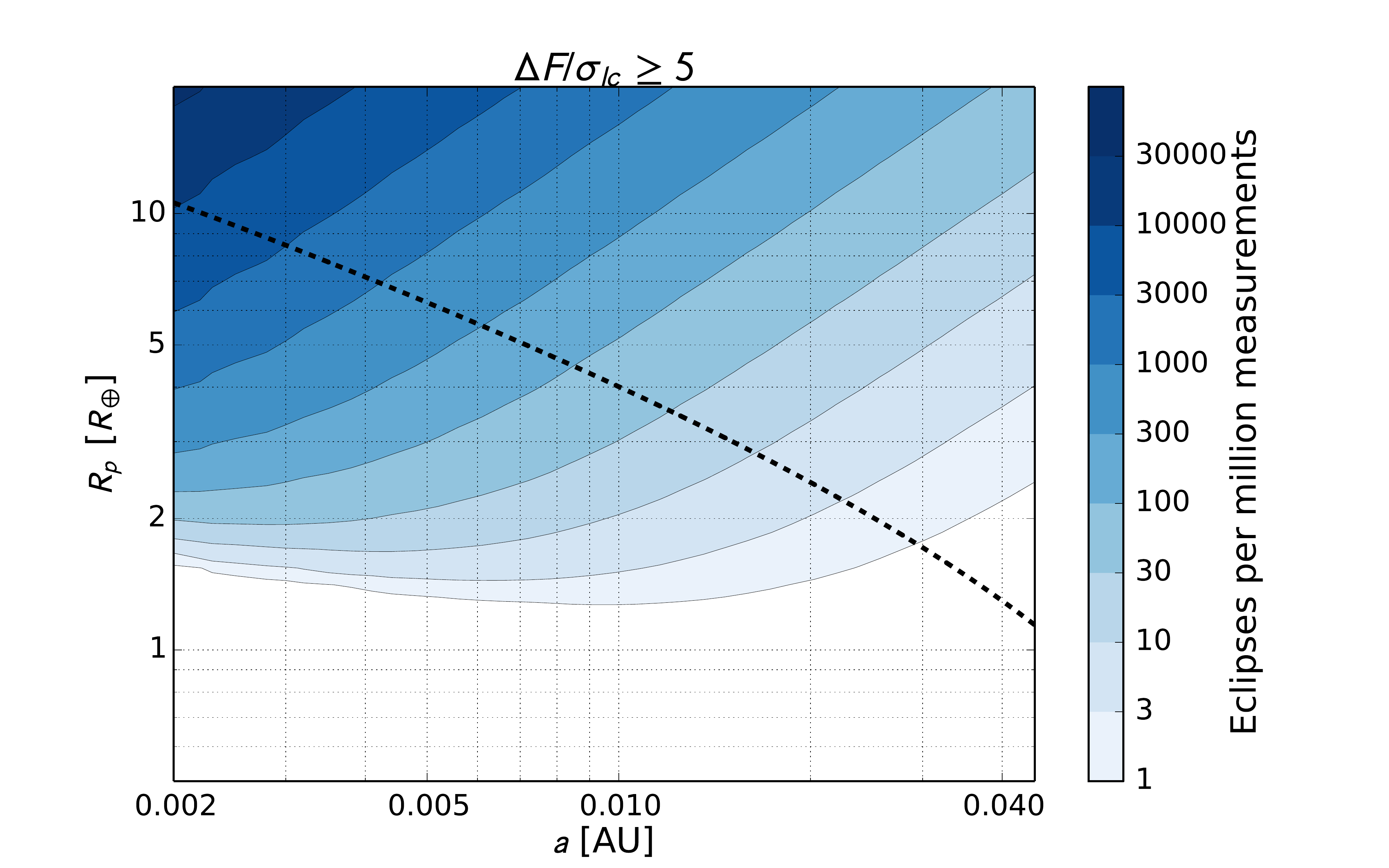}
\centering
\caption{Expected detectable eclipse rate per million exposures of the medium-deep survey. An eclipse is deemed detectable if the
depth is greater or equal to five times
the measurement uncertainty. The measurement uncertainty is calculated by adding the reported uncertainty in
quadrature with the standard deviation of the
light curve on a per filter basis. The dashed line marks the point at which the eclipse duration is equal to the 
integration time. Eclipses caused by objects
with parameters that fall in the region above and to the right of the dashed line will have eclipses that may span
multiple adjacent exposures. Our assumption
that each light curve measurement is independent is invalid in this regime and our expected eclipse rate will be slightly overestimated.
}
\label{fig:expected}
\end{figure}

For eclipses with durations shorter or equal to the exposure time the likelihood of any given measurement being in eclipse is then the sum of the probabilities for all possible orbital configurations that would produce
an observed eclipse of depth $m$. For example, a measurement with $m=0.1$ could be caused by a very small planet transiting slowly across the face
of the star with a transit duration approximately equal to the exposure time. Alternatively, an eclipse of a much larger planet causing a
complete occultation of the WD on a very short-period orbit would streak across the face of the star with a transit duration much shorter
than the exposure time. The mean flux during the exposure may look identical in these two cases. Both of these cases and all
other situations that could cause an observed eclipse depth $m$ must given the appropriate
weight in the final likelihood calculation. Figure \ref{fig:depthprob} shows the eclipse depth probability distributions for a few hypothetical scenarios.

By the definition of our eclipse detection algorithm each exposure is sensitive to eclipses of depth $m\geq5\Delta F/\sigma_{\rm lc}$. By integrating over all scenarios that would cause an observed eclipse depth greater than or equal to $5\Delta F/\sigma_{\rm lc}$ for every measurement we derive the probability that we could have detected an eclipse during each exposure if $\eta=1$.
The inverse of the summed probabilities over all exposures for all light curves gives a
total number of expected eclipses for the survey as a Poisson expectation value for the rate of eclipses (Figure \ref{fig:expected}). We then compare this Poisson distribution for the
expected number of eclipses with the lack of detected eclipses for many values of $a, p$, and $\eta$.

\subsection{Occurrence constraints}
If we treat the number of expected eclipses as a Poisson expectation value ($\lambda$), the probability that we should detect $k$ eclipses is

\begin{equation}
P(k, a, p) = \frac{\lambda(\eta,a,p)^k \exp(-\lambda(\eta,a,p))}{k!}.
\end{equation}
Since we have zero detected eclipses this can be simplified to $P(0, a, p) = \exp(-\lambda(\eta,a,p))$.
By setting $P(0,a,p)$ equal to a confidence interval $C$ and decomposing $\lambda(a,p)$ into the expectation value of eclipses if the planet occurrence rate is equal to 1 ($\lambda_1(a,p)$)
multiplied by the actual planet occurrence rate ($\eta$) we derive the maximum planet occurrence rate that is compatible with the observations at a confidence level of $C$ 
\begin{equation}
\eta \leq \frac{\ln{(1-C)}}{\lambda_1(a,p)},
\label{eqn:constraints}
\end{equation}
assuming the planet occurrence rate is constant as a function of $a$ and $p$.

\begin{figure}[h]
\epsscale{2.5}
\plottwo{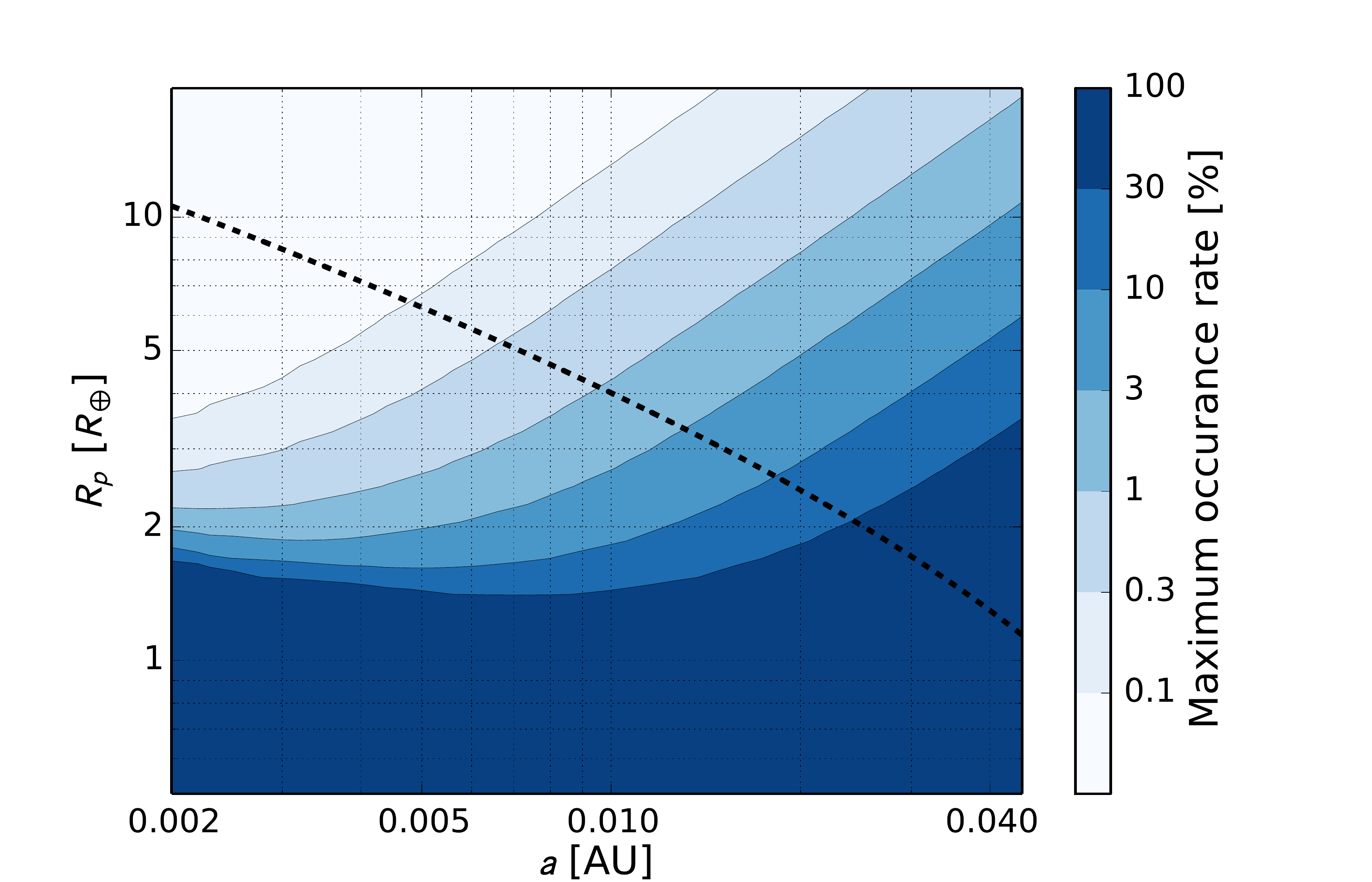}{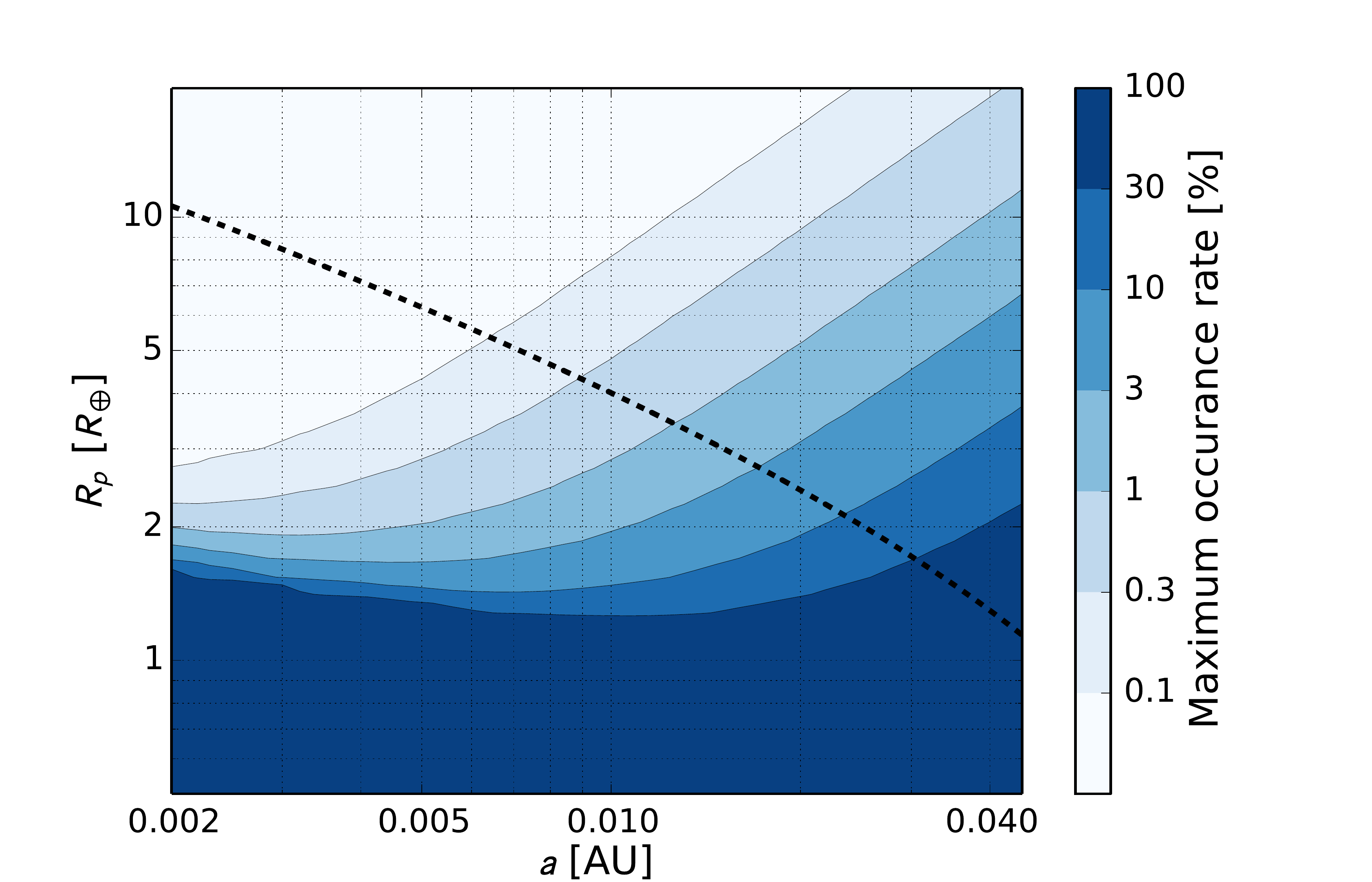}
\centering
\caption{\emph{Top:} Maximum planet occurrence rate compatible with the observations at 95\% confidence.
\emph{Bottom:} Maximum planet occurrence rate compatible with the observations at 68\% confidence.
In both panels dashed line is the same as in Figure \ref{fig:expected}. The maximum occurrence rates will be slightly underestimated in the region to the upper right of this dashed line.
}
\label{fig:constraints}
\end{figure}

\begin{figure}[h]
\epsscale{1.25}
\plotone{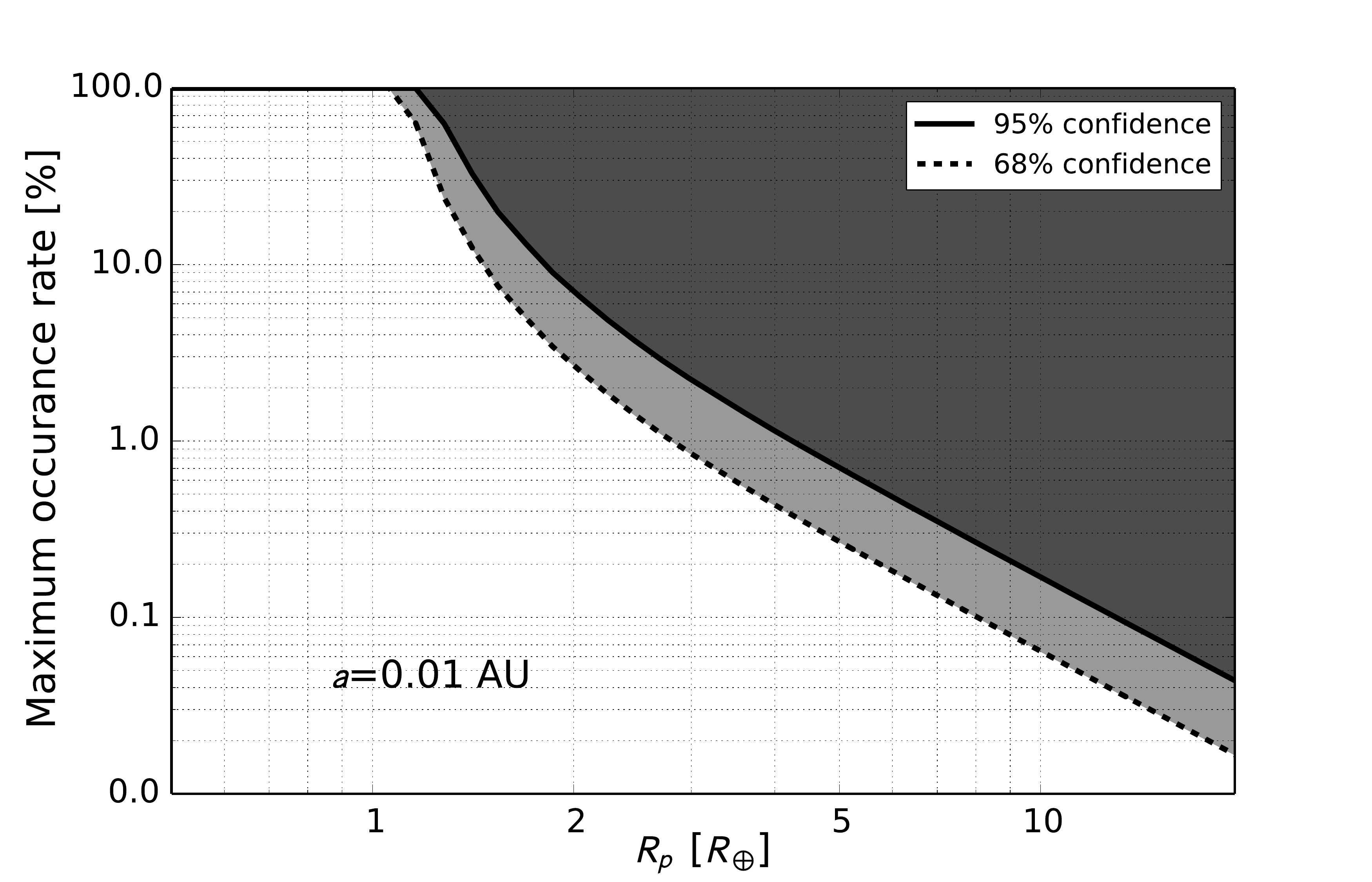}
\centering
\caption{Maximum planet occurrence rate consistent with our data as a function of planet radius at a semi-major axis of $a=0.01$ AU for confidence levels of 95\% (solid) and 68\% (dashed). Shaded regions are disfavored by our data. This plot represents a slice through Figure \ref{fig:constraints} at $a=0.01$ AU.
}
\label{fig:1Dconstraints}
\end{figure}

\begin{figure*}[ht]
\epsscale{1.15}
\plotone{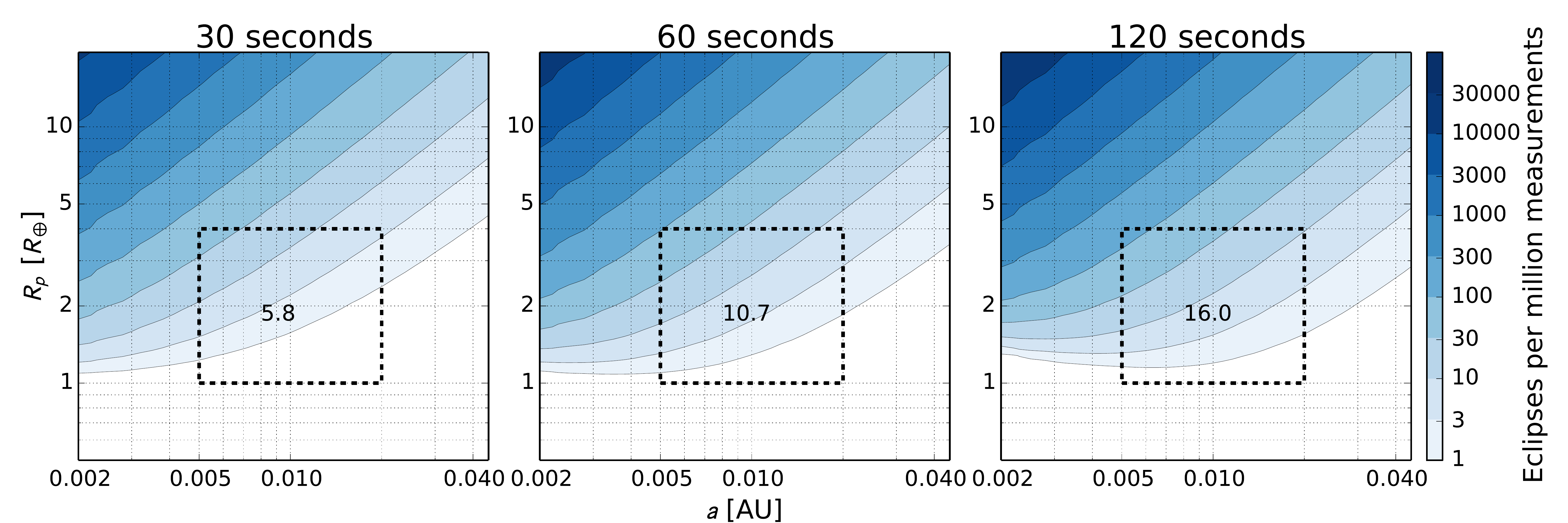}
\centering
\caption{Expected detectable eclipse rates calculated as described in \S \ref{eclipseprob} for hypothetical surveys using the \PS\-like throughput with different exposure times. The numbers within the dashed box indicate the mean eclipse rate in that region of parameter space. Shorter exposure times give increased eclipse detectability for the shortest-period objects within $\sim$0.03 AU but planets orbiting this close to their host WD would likely be ripped apart by tidal forces.
Although the mean eclipse rate in the region of interest goes up with longer exposure times this is reversed if you consider a fixed total survey exposure time (take twice as many 60 second exposures as 120 second exposures, etc.). However, the eclipse rates remain nearly constant indicating that the best way to increase sensitivity in this regime is to increase the number of epochs observed (larger number of WDs and/or higher cadence).
}
\label{fig:expect_multi}
\end{figure*}

\section{Discussion}

Although we find no convincing detections of eclipses with durations consistent with substellar objects we are still able to put strong
constraints on the WD-hosted planet occurrence rate. Figures \ref{fig:constraints} and \ref{fig:1Dconstraints} show the maximum occurrence rate that is consistent with
our observations at 95\% and 68\% confidence levels assuming R$_{\rm WD}$ = 0.01 R$_{\odot}$ and M$_{\rm WD}$ = 0.6 M$_{\odot}$. This should be a relatively good approximation since the masses and radii of most WDs fall close to these values. For each reported occurrence rate ($\eta$) we first state the value corresponding to the maximum allowable occurrence rate averaged over the specified region of interest for the 95\% confidence limit and then the 68\% confidence limit immediately following in parenthesis. For example, our results suggest that less than 0.4\% (0.2\%) of WDs host planets with radii greater than
$\sim$2 Earth radii and semi-major axis between 0.002 and 0.01 AU. 0.4\% is the maximum occurrence rate allowed by our data at 95\% confidence and 0.2\% is the same for a confidence level of 68\%.

It is an interesting exercise to break up the two-dimensional occurrence limits into regions that correspond to classes of planets that we are 
more familiar with orbiting main-sequence stars. Other studies have shown similarities between the architectures of exoplanetary systems
around low-mass M-dwarfs with the moons of Jupiter \citep{Muirhead12} and scaled-down versions of our solar system or exoplanetary systems around more massive stars. If we scale down the orbital distances of the known exoplanet population we can look at the occurrence limits in a few interesting regimes; hot Jupiters, hot super-Earths, and habitable-zone super-Earths.

The Roche limit for a fluid body with mean density $\rho_p$ orbiting a WD with density $\rho_{\rm WD}$ and radius $R_{\rm WD}$ can be approximated as 
\begin{equation}
L_{R} \approx 2.44 R_{\rm WD} \left( \frac{\rho_{\rm WD}}{\rho_p} \right)^{1/3}.
\end{equation}
For our assumed WD properties the Roche limit for a Jupiter-like planet is $L_{R}\approx0.01$ AU. It is not surprising that we do not detect
any Jupiter-sized objects inside 0.01 AU. However, we can equate a population of Jupiter-sized planets orbiting between
0.01 and 0.04 AU to the hot Jupiters observed orbiting very close to solar-type stars. In this regime an eclipse duration is slightly longer
than the duration of a single exposure. Therefore our expectation value for eclipses is slightly overestimated, however we do not expect this to be
the dominant source of error in the occurrence rate limits.
The mean maximum occurrence rate for WD-hosted hot Jupiters ($R=10-20 R_{\oplus}$) is 0.5\% (0.2\%). Indicating that hot Jupiters around WDs are
very rare or non-existent. This is in good agreement with the frequency of hot Jupiters around solar-type stars measured to be between 0.3\% and 1.5\%
\citep{Marcy05a, Gould06, Cumming08, Howard11, Mayor11, Wright12}.

A rigid body can orbit slightly closer to the WD without being tidely disrupted.
Planets with radii larger than $\sim$1.5~\Rearth\ generally have
densities lower than that of the Earth and likely have an extended gas-dominated atmosphere \citep{Weiss14}. However, some super-Earths with slightly larger
radii have high densities consistent with a rocky composition, e.g. CoRoT-7b \citep{Leger09}, Kepler-20b \citep{Gautier12}, and Kepler-19b \citep{Ballard11}.
This class of planets may be the remaining cores of evaporated gas giant planets \citep{Hebrard04}. Our results suggest that less than 1.5\% (0.6\%) of WDs host planets
with radii between 2.0 and 5.0 \Rearth\ orbiting with semi-major axis between 0.005 and 0.01 AU. \citet{Howard12} measure an occurrence rate of 13\% for 2-4 \Rearth\ planets with orbital periods shorter than 50 days. However, the occurrence rate drops with shorter orbital periods to 2.5\% for periods shorter than
10 days. Our lack of detections indicate that hot super-Earths are almost certainly less common around WDs than they are around solar-type stars.

Perhaps the most interesting planets to consider are those that have an equilibrium temperature such that they could sustain liquid water on their surfaces.
Since WDs cool and decrease in luminosity as they age, the habitable zone (HZ) boundaries also change as a function of time.
\citet{Agol11} define the WD continuous habitable zone (CHZ) as the range of semi-major axis that would be within the HZ for a minimum of 3 Gyr and also
outside of the tidal destruction radius for an Earth-density planet.
For a 0.6 M$_\odot$ WD this corresponds semi-major axis between 0.005 and 0.02 AU. Our data show that planets in the CHZ with radii between 2-5 \Rearth\ could be present around no more than 3.4\% (1.3\%) of WDs. This is significantly less than the predicted frequency of Earth-size planets in the habitable zone of solar-type stars \citep[$\sim$22\%,][]{Petigura13}.


A large population of short-period planets orbiting solar-type and
 M-dwarf stars has been observed.  We might expect WDs to host
 similar planets if they can reform from a post-giant phase debris
 disk or migrate from larger orbital distances once the star becomes
 a WD.  However, our observations are quite sensitive to planets
 larger than the Earth orbiting close to the WD, and the lack of any
 eclipses suggests that these processes are highly inefficient if
 they occur at all.  There are very few planets in short-period
 orbits around WDs.

\subsection{Future survey design}
\label{sec:future}
Since eclipse times are generally shorter than the 4 min exposure times for the medium-deep survey we explore the idea of designing a similar survey
with shorter exposure times and decreased sensitivity to shallow eclipses. This would cause less dilution of the eclipse signals over the duration of the exposure. We re-calculate the expected eclipse rates for exposure times of 30, 60, and 120 seconds scaling the measured noise properties from our 240 s data. We use the mean eclipse rate for planets with radii between 1-5 \Rearth\ orbiting between 0.005 and 0.02 AU as a metric for comparison. Figure \ref{fig:expect_multi} illustrates the result. We find that decreasing the exposure time gives a modest boost in sensitivity to these planets for a given total
survey exposure time. The most dramatic increase in sensitivity when going to short exposure times is for the very short period planets orbiting interior to 0.003 AU. However, planets are not able to withstand the tidal forces this close to the WD so we would not expect planets to exist in this regime. The expected eclipse rate in our region of interest is dominated by signal-to-noise of the individual detections. Although the eclipses are diluted by long exposure times, this is balanced by the increased gain in sensitivity to these shallow, diluted eclipses due to the greater signal to noise obtained in longer exposures. This suggests that the best way to detect these Earth to Neptune size planets in the WD CHZ may be to increase the etendue of the survey to detect more WDs on a greater number of epochs by covering a large area of the sky at high cadence. The ATLAS \citep{Tonry11} and Large Synoptic Survey Telescope \citep{Ivesic08} surveys should be ideal for detecting these extremely rare events.

\subsection{\PS\ 3$\pi$}

The \PS\ 3$\pi$ survey covers 30,000 square degrees with approximately 60 observational epochs per object \citep{Kaiser10, Magnier13}. The depth and cadence are inferior to that of the medium deep fields, but the huge amount of sky observed makes it interesting to explore the contribution that this survey could make to the occurrence rate limits if we were to perform a similar analysis on a combined dataset.

We start with an order-of magnitude estimate of the number of WDs we would expect to find in the 3$\pi$ survey data via reduced proper motion. The exposure times for the 3$\pi$ survey are 60 seconds vs. 240 seconds for the medium deep fields, but let us assume that our ability to detect WDs is limited by the length of the observational baseline and not by signal to noise of the detections. Since the sky coverage is a factor of $\sim$400 greater in the 3$\pi$ survey it is reasonable to scale the number of astrometrically-selected WDs found in the medium-deep fields (661) by 400. Therefore, we expect to find $\sim$30,000 WDs via reduced proper motion in the 3$\pi$ data. Since each WD is observed 60 times this gives a total of 1.8 million measurements. The shorter exposure times increase our sensitivity to very short duration eclipses, however the largest gain in sensitivity is to planets orbiting well inside the tidal destruction radius (see Figure \ref{fig:expect_multi} and \S \ref{sec:future}). Combining these 1.8 million epochs with the 4.3 million epochs from the medium-deep fields increases our total number of measurements by a factor of 1.4 and strengthens (decreases) our maximum occurrence constraints by this same factor. This $\sim\sqrt{2}$ improvement would not change our primary conclusion that planets around WDs are rare.

\section{Conclusions}

Our systematic search for eclipses of WDs in the \PS\ medium-deep fields places strong constraints on the WD planet occurrence rates.
We analyze a sample of $\sim$3000 WDs selected via proper motion and color along with a control sample of $\sim$1200 stars. These
WDs were observed for 5 years on over 4.3 million epochs.

We search for potential eclipses by identifying low outliers in the light curves. A total of 133 candidate eclipses are identified after applying a series
of photometry then image-based filters to remove outliers caused by weather, CCD artifacts, or an improperly modeled PSF. After visual inspection
of all candidates we find none that is consistent with an eclipse or occultation by a substellar object.

We calculate the number of expected eclipses if every WD hosted at least one planet ($\eta=1$) by convolving a trapezoidal transit model with the survey
exposure time and integrating over all possible geometric orientations and many values of $R_p$ and $a$.
The expected number of eclipses are treated as a Poisson expectation value for the rate of events
which are converted into 95\% (68\%) confidence intervals. We then invert these rates to obtain the maximum value of $\eta$ that is consistent with our data.

Our results suggest that hot Jupiters around WDs are at least as rare as they are around solar-type stars, occurring around no more than 0.5\% (0.2\%) of WDs.
Hot super-Earths occur around no more than 1.5\% (0.6\%) of stars, and super-Earths in the CHZ are present around no more than 3.4\% (1.3\%) of WDs.
All evidence presented in this study indicate that short-period planets around WDs are significantly less abundant than short-period planets orbiting main-sequence stars.

\acknowledgments{
The Pan-STARRS1 Surveys (PS1) have been made possible through contributions of the Institute for Astronomy, the University of Hawaii, the Pan-STARRS Project Office, the Max-Planck Society and its participating institutes, the Max Planck Institute for Astronomy, Heidelberg and the Max Planck Institute for Extraterrestrial Physics, Garching, The Johns Hopkins University, Durham University, the University of Edinburgh, Queen's University Belfast, the Harvard-Smithsonian Center for Astrophysics, the Las Cumbres Observatory Global Telescope Network Incorporated, the National Central University of Taiwan, the Space Telescope Science Institute, the National Aeronautics and Space Administration under Grant No. NNX08AR22G issued through the Planetary Science Division of the NASA Science Mission Directorate, the National Science Foundation under Grant No. AST-1238877, the University of Maryland, and Eotvos Lorand University (ELTE). Support for this work was provided by National Science Foundation grant AST-1009749. Finally, we would like to thank Prof. Andrew Gould for his critical review and extremely helpful suggestions that greatly enhanced the quality of this work.
}


\bibliographystyle{apj}
\bibliography{references}

\begin{thebibliography}{51}
\expandafter\ifx\csname natexlab\endcsname\relax\def\natexlab#1{#1}\fi

\bibitem[{{Agol}(2011)}]{Agol11}
{Agol}, E. 2011, \apjl, 731, L31

\bibitem[{{Alard}(2000)}]{Alard00}
{Alard}, C. 2000, \aaps, 144, 363

\bibitem[{{Bakos} {et~al.}(2004){Bakos}, {Noyes}, {Kov{\'a}cs}, {Stanek},
  {Sasselov}, \& {Domsa}}]{Bakos04}
{Bakos}, G., {Noyes}, R.~W., {Kov{\'a}cs}, G., {Stanek}, K.~Z., {Sasselov},
  D.~D., \& {Domsa}, I. 2004, \pasp, 116, 266

\bibitem[{{Ballard} {et~al.}(2011){Ballard}, {Fabrycky}, {Fressin},
  {Charbonneau}, {Desert}, {Torres}, {Marcy}, {Burke}, {Isaacson}, {Henze},
  {Steffen}, {Ciardi}, {Howell}, {Cochran}, {Endl}, {Bryson}, {Rowe}, {Holman},
  {Lissauer}, {Jenkins}, {Still}, {Ford}, {Christiansen}, {Middour}, {Haas},
  {Li}, {Hall}, {McCauliff}, {Batalha}, {Koch}, \& {Borucki}}]{Ballard11}
{Ballard}, S., {et~al.} 2011, \apj, 743, 200

\bibitem[{{Borucki} {et~al.}(2010){Borucki}, {Koch}, {Basri}, {Batalha},
  {Brown}, {Caldwell}, {Caldwell}, {Christensen-Dalsgaard}, {Cochran},
  {DeVore}, {Dunham}, {Dupree}, {Gautier}, {Geary}, {Gilliland}, {Gould},
  {Howell}, {Jenkins}, {Kondo}, {Latham}, {Marcy}, {Meibom}, {Kjeldsen},
  {Lissauer}, {Monet}, {Morrison}, {Sasselov}, {Tarter}, {Boss}, {Brownlee},
  {Owen}, {Buzasi}, {Charbonneau}, {Doyle}, {Fortney}, {Ford}, {Holman},
  {Seager}, {Steffen}, {Welsh}, {Rowe}, {Anderson}, {Buchhave}, {Ciardi},
  {Walkowicz}, {Sherry}, {Horch}, {Isaacson}, {Everett}, {Fischer}, {Torres},
  {Johnson}, {Endl}, {MacQueen}, {Bryson}, {Dotson}, {Haas}, {Kolodziejczak},
  {Van Cleve}, {Chandrasekaran}, {Twicken}, {Quintana}, {Clarke}, {Allen},
  {Li}, {Wu}, {Tenenbaum}, {Verner}, {Bruhweiler}, {Barnes}, \&
  {Prsa}}]{Borucki10}
{Borucki}, W.~J., {et~al.} 2010, Science, 327, 977

\bibitem[{{Burleigh} {et~al.}(2006){Burleigh}, {Hogan}, {Dobbie}, {Napiwotzki},
  \& {Maxted}}]{Burleigh06}
{Burleigh}, M.~R., {Hogan}, E., {Dobbie}, P.~D., {Napiwotzki}, R., \& {Maxted},
  P.~F.~L. 2006, \mnras, 373, L55

\bibitem[{{Cumming} {et~al.}(2008){Cumming}, {Butler}, {Marcy}, {Vogt},
  {Wright}, \& {Fischer}}]{Cumming08}
{Cumming}, A., {Butler}, R.~P., {Marcy}, G.~W., {Vogt}, S.~S., {Wright}, J.~T.,
  \& {Fischer}, D.~A. 2008, \pasp, 120, 531

\bibitem[{{Debes} {et~al.}(2011){Debes}, {Hoard}, {Kilic}, {Wachter},
  {Leisawitz}, {Cohen}, {Kirkpatrick}, \& {Griffith}}]{Debes11}
{Debes}, J.~H., {Hoard}, D.~W., {Kilic}, M., {Wachter}, S., {Leisawitz}, D.~T.,
  {Cohen}, M., {Kirkpatrick}, J.~D., \& {Griffith}, R.~L. 2011, \apj, 729, 4

\bibitem[{{Debes} {et~al.}(2005){Debes}, {Sigurdsson}, \&
  {Woodgate}}]{Debes05a}
{Debes}, J.~H., {Sigurdsson}, S., \& {Woodgate}, B.~E. 2005, \aj, 130, 1221

\bibitem[{{Devor}(2005)}]{Devor05}
{Devor}, J. 2005, \apj, 628, 411

\bibitem[{{Drake} {et~al.}(2010){Drake}, {Beshore}, {Catelan}, {Djorgovski},
  {Graham}, {Kleinman}, {Larson}, {Mahabal}, \& {Williams}}]{Drake10}
{Drake}, A.~J., {et~al.} 2010, ArXiv e-prints

\bibitem[{{Faedi} {et~al.}(2011){Faedi}, {West}, {Burleigh}, {Goad}, \&
  {Hebb}}]{Faedi11}
{Faedi}, F., {West}, R.~G., {Burleigh}, M.~R., {Goad}, M.~R., \& {Hebb}, L.
  2011, \mnras, 410, 899

\bibitem[{{Farihi} {et~al.}(2008){Farihi}, {Becklin}, \&
  {Zuckerman}}]{Farihi08}
{Farihi}, J., {Becklin}, E.~E., \& {Zuckerman}, B. 2008, \apj, 681, 1470

\bibitem[{{Farihi} {et~al.}(2005){Farihi}, {Zuckerman}, \&
  {Becklin}}]{Farihi05}
{Farihi}, J., {Zuckerman}, B., \& {Becklin}, E.~E. 2005, \aj, 130, 2237

\bibitem[{{Gaudi}(2012)}]{Gaudi12}
{Gaudi}, B.~S. 2012, \araa, 50, 411

\bibitem[{{Gautier} {et~al.}(2012){Gautier}, {Charbonneau}, {Rowe}, {Marcy},
  {Isaacson}, {Torres}, {Fressin}, {Rogers}, {D{\'e}sert}, {Buchhave},
  {Latham}, {Quinn}, {Ciardi}, {Fabrycky}, {Ford}, {Gilliland}, {Walkowicz},
  {Bryson}, {Cochran}, {Endl}, {Fischer}, {Howell}, {Horch}, {Barclay},
  {Batalha}, {Borucki}, {Christiansen}, {Geary}, {Henze}, {Holman}, {Ibrahim},
  {Jenkins}, {Kinemuchi}, {Koch}, {Lissauer}, {Sanderfer}, {Sasselov},
  {Seager}, {Silverio}, {Smith}, {Still}, {Stumpe}, {Tenenbaum}, \& {Van
  Cleve}}]{Gautier12}
{Gautier}, III, T.~N., {et~al.} 2012, \apj, 749, 15

\bibitem[{{Gould} {et~al.}(2006){Gould}, {Udalski}, {An}, {Bennett}, {Zhou},
  {Dong}, {Rattenbury}, {Gaudi}, {Yock}, {Bond}, {Christie}, {Horne},
  {Anderson}, {Stanek}, {DePoy}, {Han}, {McCormick}, {Park}, {Pogge},
  {Poindexter}, {Soszy{\'n}ski}, {Szyma{\'n}ski}, {Kubiak}, {Pietrzy{\'n}ski},
  {Szewczyk}, {Wyrzykowski}, {Ulaczyk}, {Paczy{\'n}ski}, {Bramich},
  {Snodgrass}, {Steele}, {Burgdorf}, {Bode}, {Botzler}, {Mao}, \&
  {Swaving}}]{Gould06}
{Gould}, A., {et~al.} 2006, \apjl, 644, L37

\bibitem[{{Hartman} {et~al.}(2008){Hartman}, {Gaudi}, {Holman}, {McLeod},
  {Stanek}, {Barranco}, {Pinsonneault}, {Meibom}, \& {Kalirai}}]{Hartman08}
{Hartman}, J.~D., {et~al.} 2008, \apj, 675, 1233

\bibitem[{{H{\'e}brard} {et~al.}(2004){H{\'e}brard}, {Lecavelier Des
  {\'E}tangs}, {Vidal-Madjar}, {D{\'e}sert}, \& {Ferlet}}]{Hebrard04}
{H{\'e}brard}, G., {Lecavelier Des {\'E}tangs}, A., {Vidal-Madjar}, A.,
  {D{\'e}sert}, J.-M., \& {Ferlet}, R. 2004, in Astronomical Society of the
  Pacific Conference Series, Vol. 321, Extrasolar Planets: Today and Tomorrow,
  ed. J.~{Beaulieu}, A.~{Lecavelier Des Etangs}, \& C.~{Terquem}, 203

\bibitem[{{Hogan} {et~al.}(2009){Hogan}, {Burleigh}, \& {Clarke}}]{Hogan09}
{Hogan}, E., {Burleigh}, M.~R., \& {Clarke}, F.~J. 2009, \mnras, 396, 2074

\bibitem[{{Howard} {et~al.}(2010){Howard}, {Johnson}, {Marcy}, {Fischer},
  {Wright}, {Bernat}, {Henry}, {Peek}, {Isaacson}, {Apps}, {Endl}, {Cochran},
  {Valenti}, {Anderson}, \& {Piskunov}}]{Howard10b}
{Howard}, A.~W., {et~al.} 2010, \apj, 721, 1467

\bibitem[{Howard {et~al.}(2010)Howard, Marcy, Johnson, Fischer, Wright,
  Isaacson, Valenti, Anderson, Lin, \& Ida}]{Howard10a}
Howard, A.~W., {et~al.} 2010, Science, 330, 653

\bibitem[{{Howard} {et~al.}(2011){Howard}, {Johnson}, {Marcy}, {Fischer},
  {Wright}, {Henry}, {Isaacson}, {Valenti}, {Anderson}, \&
  {Piskunov}}]{Howard11}
{Howard}, A.~W., {et~al.} 2011, \apj, 726, 73

\bibitem[{{Howard} {et~al.}(2012){Howard}, {Marcy}, {Bryson}, {Jenkins},
  {Rowe}, {Batalha}, {Borucki}, {Koch}, {Dunham}, {Gautier}, {Van Cleve},
  {Cochran}, {Latham}, {Lissauer}, {Torres}, {Brown}, {Gilliland}, {Buchhave},
  {Caldwell}, {Christensen-Dalsgaard}, {Ciardi}, {Fressin}, {Haas}, {Howell},
  {Kjeldsen}, {Seager}, {Rogers}, {Sasselov}, {Steffen}, {Basri},
  {Charbonneau}, {Christiansen}, {Clarke}, {Dupree}, {Fabrycky}, {Fischer},
  {Ford}, {Fortney}, {Tarter}, {Girouard}, {Holman}, {Johnson}, {Klaus},
  {Machalek}, {Moorhead}, {Morehead}, {Ragozzine}, {Tenenbaum}, {Twicken},
  {Quinn}, {Isaacson}, {Shporer}, {Lucas}, {Walkowicz}, {Welsh}, {Boss},
  {Devore}, {Gould}, {Smith}, {Morris}, {Prsa}, {Morton}, {Still}, {Thompson},
  {Mullally}, {Endl}, \& {MacQueen}}]{Howard12}
---. 2012, \apjs, 201, 15

\bibitem[{{Ivezic} {et~al.}(2008){Ivezic}, {Tyson}, {Acosta}, {Allsman},
  {Anderson}, {Andrew}, {Angel}, {Axelrod}, {Barr}, {Becker}, {Becla},
  {Beldica}, {Blandford}, {Bloom}, {Borne}, {Brandt}, {Brown}, {Bullock},
  {Burke}, {Chandrasekharan}, {Chesley}, {Claver}, {Connolly}, {Cook},
  {Cooray}, {Covey}, {Cribbs}, {Cutri}, {Daues}, {Delgado}, {Ferguson},
  {Gawiser}, {Geary}, {Gee}, {Geha}, {Gibson}, {Gilmore}, {Gressler}, {Hogan},
  {Huffer}, {Jacoby}, {Jain}, {Jernigan}, {Jones}, {Juric}, {Kahn}, {Kalirai},
  {Kantor}, {Kessler}, {Kirkby}, {Knox}, {Krabbendam}, {Krughoff}, {Kulkarni},
  {Lambert}, {Levine}, {Liang}, {Lim}, {Lupton}, {Marshall}, {Marshall}, {May},
  {Miller}, {Mills}, {Monet}, {Neill}, {Nordby}, {O'Connor}, {Oliver},
  {Olivier}, {Olsen}, {Owen}, {Peterson}, {Petry}, {Pierfederici},
  {Pietrowicz}, {Pike}, {Pinto}, {Plante}, {Radeka}, {Rasmussen}, {Ridgway},
  {Rosing}, {Saha}, {Schalk}, {Schindler}, {Schneider}, {Schumacher}, {Sebag},
  {Seppala}, {Shipsey}, {Silvestri}, {Smith}, {Smith}, {Strauss}, {Stubbs},
  {Sweeney}, {Szalay}, {Thaler}, {Vanden Berk}, {Walkowicz}, {Warner},
  {Willman}, {Wittman}, {Wolff}, {Wood-Vasey}, {Yoachim}, {Zhan}, \& {for the
  LSST Collaboration}}]{Ivesic08}
{Ivezic}, Z., {et~al.} 2008, ArXiv e-prints

\bibitem[{{Johnson} {et~al.}(2007){Johnson}, {Fischer}, {Marcy}, {Wright},
  {Driscoll}, {Butler}, {Hekker}, {Reffert}, \& {Vogt}}]{Johnson07}
{Johnson}, J.~A., {et~al.} 2007, \apj, 665, 785

\bibitem[{{Jura} {et~al.}(2009){Jura}, {Farihi}, \& {Zuckerman}}]{Jura09}
{Jura}, M., {Farihi}, J., \& {Zuckerman}, B. 2009, \aj, 137, 3191

\bibitem[{{Kaiser} {et~al.}(2010){Kaiser}, {Burgett}, {Chambers}, {Denneau},
  {Heasley}, {Jedicke}, {Magnier}, {Morgan}, {Onaka}, \& {Tonry}}]{Kaiser10}
{Kaiser}, N., {et~al.} 2010, in Society of Photo-Optical Instrumentation
  Engineers (SPIE) Conference Series, Vol. 7733, Society of Photo-Optical
  Instrumentation Engineers (SPIE) Conference Series

\bibitem[{{Kilic} {et~al.}(2009){Kilic}, {Gould}, \& {Koester}}]{Kilic09}
{Kilic}, M., {Gould}, A., \& {Koester}, D. 2009, \apj, 705, 1219

\bibitem[{{Kov{\'a}cs} {et~al.}(2002){Kov{\'a}cs}, {Zucker}, \&
  {Mazeh}}]{Kovacs02}
{Kov{\'a}cs}, G., {Zucker}, S., \& {Mazeh}, T. 2002, \aap, 391, 369

\bibitem[{{L{\'e}ger} {et~al.}(2009){L{\'e}ger}, {Rouan}, {Schneider}, {Barge},
  {Fridlund}, {Samuel}, {Ollivier}, {Guenther}, {Deleuil}, {Deeg}, {Auvergne},
  {Alonso}, {Aigrain}, {Alapini}, {Almenara}, {Baglin}, {Barbieri}, {Bruntt},
  {Bord{\'e}}, {Bouchy}, {Cabrera}, {Catala}, {Carone}, {Carpano}, {Csizmadia},
  {Dvorak}, {Erikson}, {Ferraz-Mello}, {Foing}, {Fressin}, {Gandolfi},
  {Gillon}, {Gondoin}, {Grasset}, {Guillot}, {Hatzes}, {H{\'e}brard}, {Jorda},
  {Lammer}, {Llebaria}, {Loeillet}, {Mayor}, {Mazeh}, {Moutou}, {P{\"a}tzold},
  {Pont}, {Queloz}, {Rauer}, {Renner}, {Samadi}, {Shporer}, {Sotin}, {Tingley},
  {Wuchterl}, {Adda}, {Agogu}, {Appourchaux}, {Ballans}, {Baron}, {Beaufort},
  {Bellenger}, {Berlin}, {Bernardi}, {Blouin}, {Baudin}, {Bodin}, {Boisnard},
  {Boit}, {Bonneau}, {Borzeix}, {Briet}, {Buey}, {Butler}, {Cailleau},
  {Cautain}, {Chabaud}, {Chaintreuil}, {Chiavassa}, {Costes}, {Cuna Parrho},
  {de Oliveira Fialho}, {Decaudin}, {Defise}, {Djalal}, {Epstein}, {Exil},
  {Faur{\'e}}, {Fenouillet}, {Gaboriaud}, {Gallic}, {Gamet}, {Gavalda},
  {Grolleau}, {Gruneisen}, {Gueguen}, {Guis}, {Guivarc'h}, {Guterman},
  {Hallouard}, {Hasiba}, {Heuripeau}, {Huntzinger}, {Hustaix}, {Imad},
  {Imbert}, {Johlander}, {Jouret}, {Journoud}, {Karioty}, {Kerjean},
  {Lafaille}, {Lafond}, {Lam-Trong}, {Landiech}, {Lapeyrere}, {Larqu{\'e}},
  {Laudet}, {Lautier}, {Lecann}, {Lefevre}, {Leruyet}, {Levacher}, {Magnan},
  {Mazy}, {Mertens}, {Mesnager}, {Meunier}, {Michel}, {Monjoin}, {Naudet},
  {Nguyen-Kim}, {Orcesi}, {Ottacher}, {Perez}, {Peter}, {Plasson}, {Plesseria},
  {Pontet}, {Pradines}, {Quentin}, {Reynaud}, {Rolland}, {Rollenhagen},
  {Romagnan}, {Russ}, {Schmidt}, {Schwartz}, {Sebbag}, {Sedes}, {Smit},
  {Steller}, {Sunter}, {Surace}, {Tello}, {Tiph{\`e}ne}, {Toulouse}, {Ulmer},
  {Vandermarcq}, {Vergnault}, {Vuillemin}, \& {Zanatta}}]{Leger09}
{L{\'e}ger}, A., {et~al.} 2009, \aap, 506, 287

\bibitem[{{Magnier} {et~al.}(2013){Magnier}, {Schlafly}, {Finkbeiner}, {Juric},
  {Tonry}, {Burgett}, {Chambers}, {Flewelling}, {Kaiser}, {Kudritzki},
  {Morgan}, {Price}, {Sweeney}, \& {Stubbs}}]{Magnier13}
{Magnier}, E.~A., {et~al.} 2013, \apjs, 205, 20

\bibitem[{{Mandel} \& {Agol}(2002)}]{Mandel02}
{Mandel}, K., \& {Agol}, E. 2002, \apjl, 580, L171

\bibitem[{{Marcy} {et~al.}(2005){Marcy}, {Butler}, {Fischer}, {Vogt}, {Wright},
  {Tinney}, \& {Jones}}]{Marcy05a}
{Marcy}, G., {Butler}, R.~P., {Fischer}, D., {Vogt}, S., {Wright}, J.~T.,
  {Tinney}, C.~G., \& {Jones}, H.~R.~A. 2005, Progress of Theoretical Physics
  Supplement, 158, 24

\bibitem[{Mayor {et~al.}(2011)Mayor, Marmier, Lovis, Udry, S{\'e}gransan, Pepe,
  Benz, Bertaux, Bouchy, Dumusque, Lo~Curto, Mordasini, Queloz, \&
  Santos}]{Mayor11}
Mayor, M., {et~al.} 2011, arXiv.org, 1109, 2497

\bibitem[{{Muirhead} {et~al.}(2012){Muirhead}, {Johnson}, {Apps}, {Carter},
  {Morton}, {Fabrycky}, {Pineda}, {Bottom}, {Rojas-Ayala}, {Schlawin},
  {Hamren}, {Covey}, {Crepp}, {Stassun}, {Pepper}, {Hebb}, {Kirby}, {Howard},
  {Isaacson}, {Marcy}, {Levitan}, {Diaz-Santos}, {Armus}, \&
  {Lloyd}}]{Muirhead12}
{Muirhead}, P.~S., {et~al.} 2012, \apj, 747, 144

\bibitem[{{Mullally}(2007)}]{Mullally07}
{Mullally}, F.~R. 2007, PhD thesis, The University of Texas at Austin

\bibitem[{{Nutzman} \& {Charbonneau}(2008)}]{Nutzman08}
{Nutzman}, P., \& {Charbonneau}, D. 2008, \pasp, 120, 317

\bibitem[{{Petigura} {et~al.}(2013){Petigura}, {Howard}, \&
  {Marcy}}]{Petigura13}
{Petigura}, E.~A., {Howard}, A.~W., \& {Marcy}, G.~W. 2013, Proceedings of the
  National Academy of Science, 110, 19273

\bibitem[{{Pollacco} {et~al.}(2006){Pollacco}, {Skillen}, {Collier Cameron},
  {Christian}, {Hellier}, {Irwin}, {Lister}, {Street}, {West}, {Anderson},
  {Clarkson}, {Deeg}, {Enoch}, {Evans}, {Fitzsimmons}, {Haswell}, {Hodgkin},
  {Horne}, {Kane}, {Keenan}, {Maxted}, {Norton}, {Osborne}, {Parley}, {Ryans},
  {Smalley}, {Wheatley}, \& {Wilson}}]{Pollacco06}
{Pollacco}, D.~L., {et~al.} 2006, \pasp, 118, 1407

\bibitem[{{Robin} {et~al.}(2003){Robin}, {Reyl{\'e}}, {Derri{\`e}re}, \&
  {Picaud}}]{Robin03}
{Robin}, A.~C., {Reyl{\'e}}, C., {Derri{\`e}re}, S., \& {Picaud}, S. 2003,
  \aap, 409, 523

\bibitem[{{Schlafly} {et~al.}(2012){Schlafly}, {Finkbeiner}, {Juri{\'c}},
  {Magnier}, {Burgett}, {Chambers}, {Grav}, {Hodapp}, {Kaiser}, {Kudritzki},
  {Martin}, {Morgan}, {Price}, {Rix}, {Stubbs}, {Tonry}, \&
  {Wainscoat}}]{Schlafly12}
{Schlafly}, E.~F., {et~al.} 2012, \apj, 756, 158

\bibitem[{{Schwarzenberg-Czerny}(1989)}]{Schwarzenberg-Czerny89}
{Schwarzenberg-Czerny}, A. 1989, \mnras, 241, 153

\bibitem[{{Steele} {et~al.}(2009){Steele}, {Burleigh}, {Farihi},
  {G{\"a}nsicke}, {Jameson}, {Dobbie}, \& {Barstow}}]{Steele09}
{Steele}, P.~R., {Burleigh}, M.~R., {Farihi}, J., {G{\"a}nsicke}, B.~T.,
  {Jameson}, R.~F., {Dobbie}, P.~D., \& {Barstow}, M.~A. 2009, \aap, 500, 1207

\bibitem[{{Tonry}(2011)}]{Tonry11}
{Tonry}, J.~L. 2011, \pasp, 123, 58

\bibitem[{{Tonry} {et~al.}(2012){Tonry}, {Stubbs}, {Kilic}, {Flewelling},
  {Deacon}, {Chornock}, {Berger}, {Burgett}, {Chambers}, {Kaiser}, {Kudritzki},
  {Hodapp}, {Magnier}, {Morgan}, {Price}, \& {Wainscoat}}]{Tonry12}
{Tonry}, J.~L., {et~al.} 2012, \apj, 745, 42

\bibitem[{{Weiss} \& {Marcy}(2014)}]{Weiss14}
{Weiss}, L.~M., \& {Marcy}, G.~W. 2014, \apjl, 783, L6

\bibitem[{{Wright} {et~al.}(2012){Wright}, {Marcy}, {Howard}, {Johnson},
  {Morton}, \& {Fischer}}]{Wright12}
{Wright}, J.~T., {Marcy}, G.~W., {Howard}, A.~W., {Johnson}, J.~A., {Morton},
  T.~D., \& {Fischer}, D.~A. 2012, \apj, 753, 160

\bibitem[{{Xu} {et~al.}(2014){Xu}, {Jura}, {Koester}, {Klein}, \&
  {Zuckerman}}]{Xu14}
{Xu}, S., {Jura}, M., {Koester}, D., {Klein}, B., \& {Zuckerman}, B. 2014,
  \apj, 783, 79

\bibitem[{{Zuckerman} \& {Becklin}(1992)}]{Zuckerman92}
{Zuckerman}, B., \& {Becklin}, E.~E. 1992, \apj, 386, 260

\bibitem[{{Zuckerman} {et~al.}(2010){Zuckerman}, {Melis}, {Klein}, {Koester},
  \& {Jura}}]{Zuckerman10}
{Zuckerman}, B., {Melis}, C., {Klein}, B., {Koester}, D., \& {Jura}, M. 2010,
  \apj, 722, 725

\end{thebibliography}

\end{document}